\documentclass[prb,amsmath,amssymb,floatfix,citeautoscript]{revtex4}
\usepackage{bm}
\usepackage{epsfig}
\usepackage{hyperref}

\def\be{\begin{equation}}
\def\ee{\end{equation}}
\def\bea{\begin{eqnarray}}
\def\eea{\end{eqnarray}}

\newcommand{\alphap}{\delta}

\begin{document}

\title{Observing the origin of superconductivity in quantum critical metals}

\author{J.-H. She}
\author{B. J. Overbosch}
\author{Y.-W. Sun}
\author{Y. Liu}
\author{K. E. Schalm}
\author{J. A. Mydosh}
\author{J. Zaanen}
\affiliation{Instituut-Lorentz for Theoretical Physics, Universiteit Leiden, P. O. Box 9506, 2300 R A Leiden, The Netherlands}

\begin{abstract}
   Despite intense efforts during the last 25 years, the physics of unconventional superconductors, 
including the cuprates with a very high transition temperature, is still a 
controversial subject.
It is 
believed that superconductivity in many of these strongly correlated metallic systems 
originates
in the physics of 
quantum phase transitions, but quite diverse perspectives have emerged on the
fundamentals of the electron-pairing physics, ranging from Hertz style  critical spin fluctuation glue 
to the holographic superconductivity of string theory. 
Here we demonstrate that the gross 
energy scaling differences
that are behind these various pairing mechanisms are directly encoded
in the frequency and temperature dependence of the dynamical pair susceptibility.  This quantity can be measured 
directly via the second order Josephson effect  
and it should be possible employing modern experimental
techniques to build a `pairing telescope'  that gives a direct view on the origin of quantum critical superconductivity.
\end{abstract}

\date{\today \ [file: \jobname]}
 \maketitle

\section{Introduction and Summary}

The large variety of superconductors that are not explained by the classic
Bardeen-Cooper-Schrieffer (BCS) theory include the cuprates \cite{Norman11,Zaanen11} and iron pnictides \cite{DHLee11} with their (very) high transition
temperatures ($T_c$'s), but also the large family of low $T_c$  heavy fermion superconductors \cite{Norman11,Pfleiderer09}. These materials have in common 
that the dominance of 
electronic repulsions 
 create an environment that is {\em a priori} very unfavorable for conventional superconductivity. Their unconventional (non-$s$-wave) order parameters indeed signal that dissimilar
 physics is at work. Based on a multitude of experiments, a widely held hypothesis has arisen that the physics of many of these systems is controlled by a quantum phase transition \cite{Sachdev99,Lohneysen07,Si08,Sachdev11}. This would generate a scale invariant quantum physics in the electron system, as it does for any other second order phase transition, and  the imprint of this universal critical behavior on the metallic state creates 
the conditions for unconventional superconductivity.

We propose to test this hypothesis of quantum-criticality as the fundamental
physics underlying the onset of superconductivity directly. A clean probe can be identified: a measurement of the dynamical order-parameter susceptibility --- the Cooper pair susceptibility --- of the quantum critical superconductor in its normal state
in a large temperature and energy interval.  Four differing
theoretical views of electron-quantum-criticality that are available
--- including two brand new paradigms descending from string theory
--- all allow for explicit computations of the susceptibility
\cite{MoonChubukov10,She09,Hartnoll08,Horowitz10}. At the same time, 
the pair susceptibility can be measured directly via the so-called second order Josephson effect in superconductor-insulator-superconductor (SIS) junctions involving superconductors with different 
transition temperatures.\cite{Ferrell69,Scalapino70}

Goldman and collaborators delivered proof of principle in the 1970s by measuring the pair susceptibility in the normal state of aluminum in an aluminum-aluminum oxide-lead junction \cite{Goldman70,Goldman06}.  
In this experiment
the order parameter of the ``strong'' superconductor with a ``high'' $T^\text{high}_c$ acts as an external perturbing field on the 
metallic electron system realized above the transition of the superconductor with a much lower $T_c^\text{low}$. In the temperature regime 
$T_c^\text{low} \le  T \ll T^\text{high}_c$ and for an applied bias $eV$ less than the gap $\Delta_\text{high}$ of the strong superconductor the current through a tunneling 
junction between the two is directly proportional to the imaginary part of the  dynamical pair susceptibility. 
This higher order Cooper pair tunneling process is a second order Josephson effect: if at low temperatures the regular dc Josephson effect can be observed (i.e., a finite supercurrent at zero bias in SIS configuration), then the higher order tunneling Cooper pair process is likely to occur in the superconductor-insulator-normal-state (SIN) configuration at finite bias.  

Quite recently 
Bergeal et al. \cite{Bergeal08} succeeded to get a signal on a 60 K underdoped 
 cuprate superconductor using a 90 K cuprate source.  This was motivated by the prediction that 
an asymmetric relaxational peak would be found signaling the dominance of phase fluctuations in the order parameter dynamics of the underdoped cuprate \cite{Levin99}. Although this prediction was
 not borne out by the experiment it is for the present purposes
 quite significant that Bergeal et al. managed to isolate the second order Josephson current
 at such a high temperature (60 K) in $d$-wave superconductors where the masking effects of the quasiparticle currents should be particularly severe.
 As we will explain from our theoretical predictions, the unambiguous information regarding the quantum critical pairing mechanism resides in the large dynamical 
 range in temperature and frequency of the pair susceptibility
 meaning that in principle one should measure 
 up to temperatures of order $50 \times T_c$ and energies greater than
 ten times the gap of the weak superconductor (we set $T_c^\text{low}=T_c$
 from here on). The system that is interrogated 
 should therefore be a quantum critical system with a low $T_c$  and the natural candidates are heavy fermion superconductors characterized
 by quantum critical points at ambient conditions. We shall propose two explicit experimental approaches using modern thin film techniques and STM/STS/PCS techniques with a superconducting tip to obtain the pair-susceptibility in the range of temperature and frequency that will distinguish between the differing quantum-critical metal models.

\begin{figure*}
\includegraphics[width=2cm]{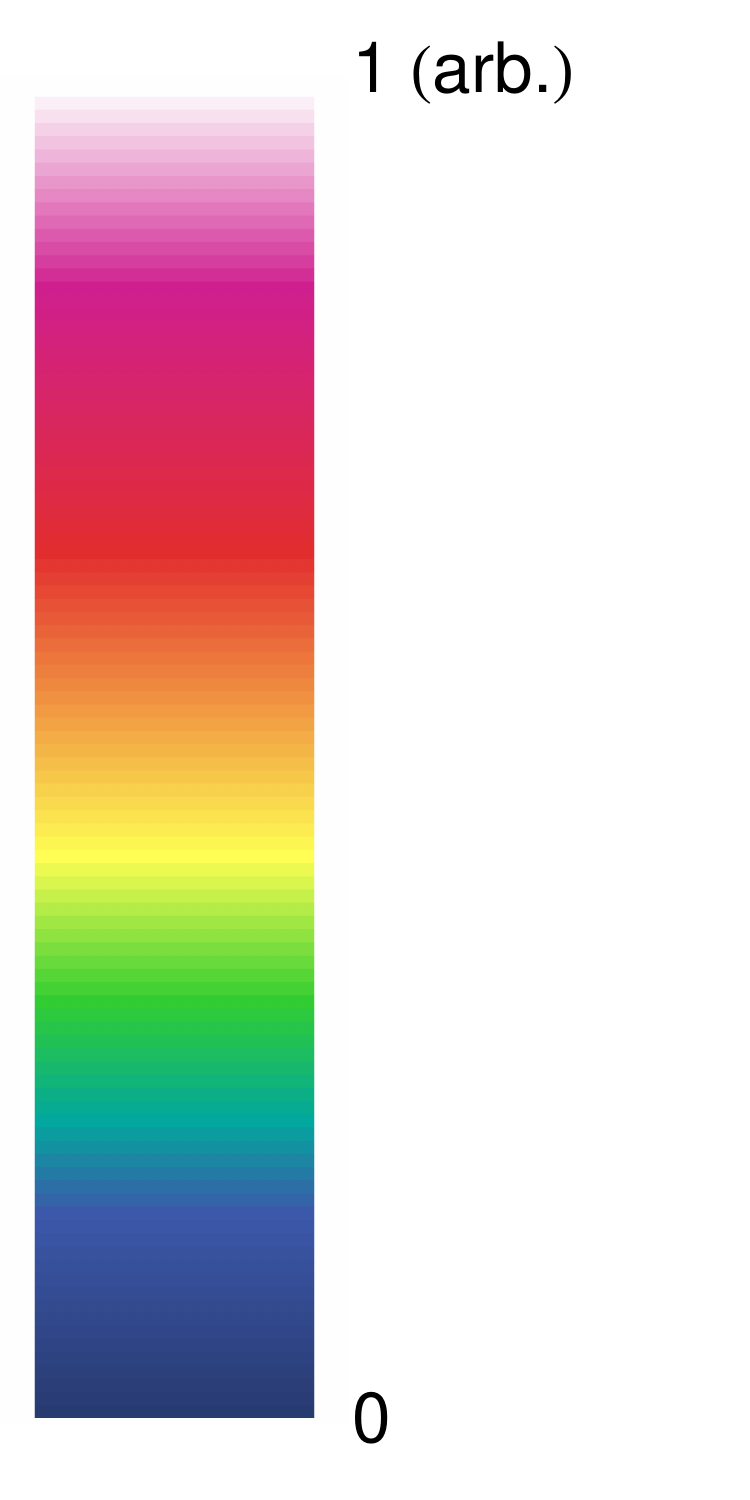}
\includegraphics[width=0.32\textwidth]{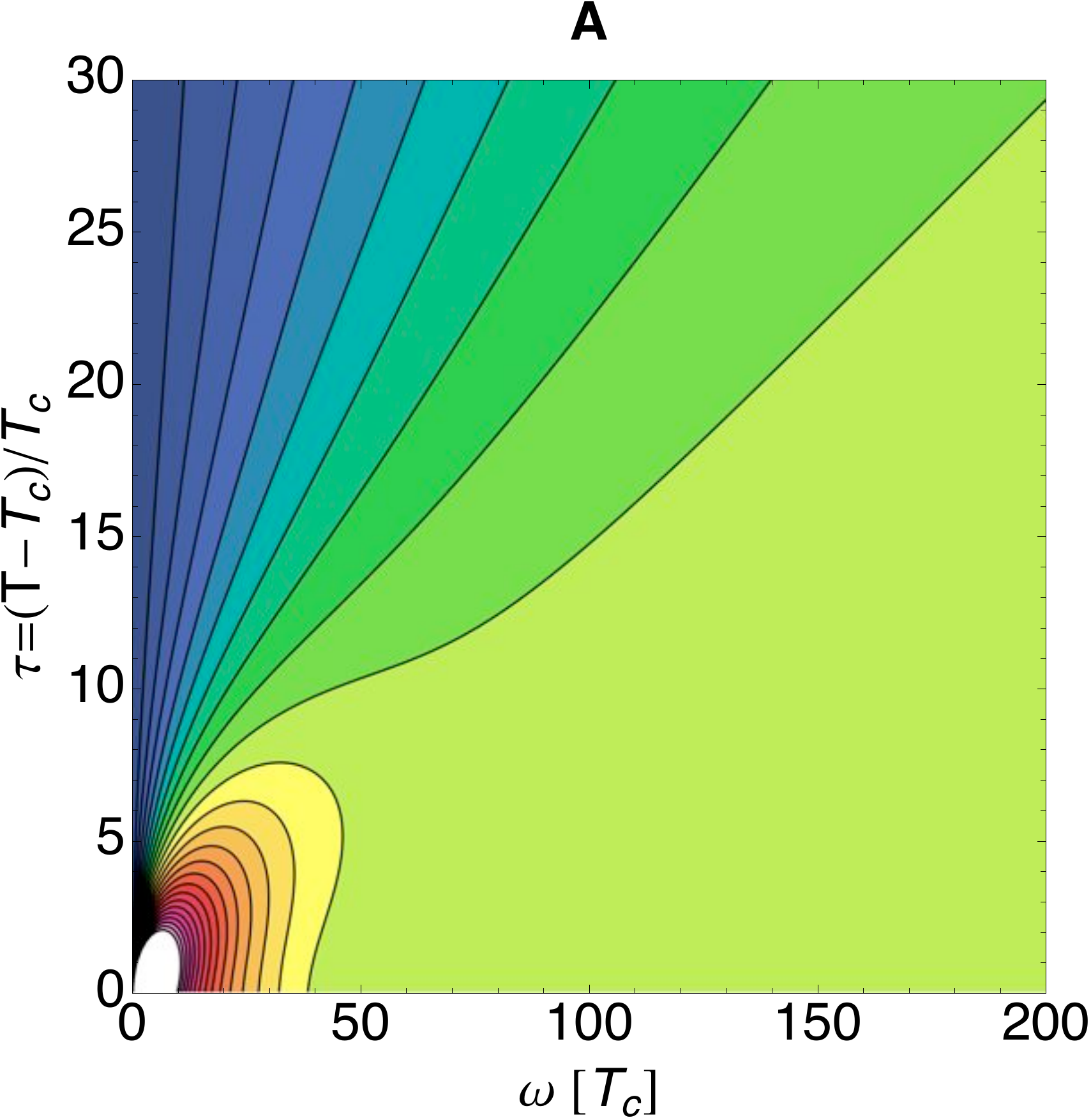}\quad
\includegraphics[width=0.32\textwidth]{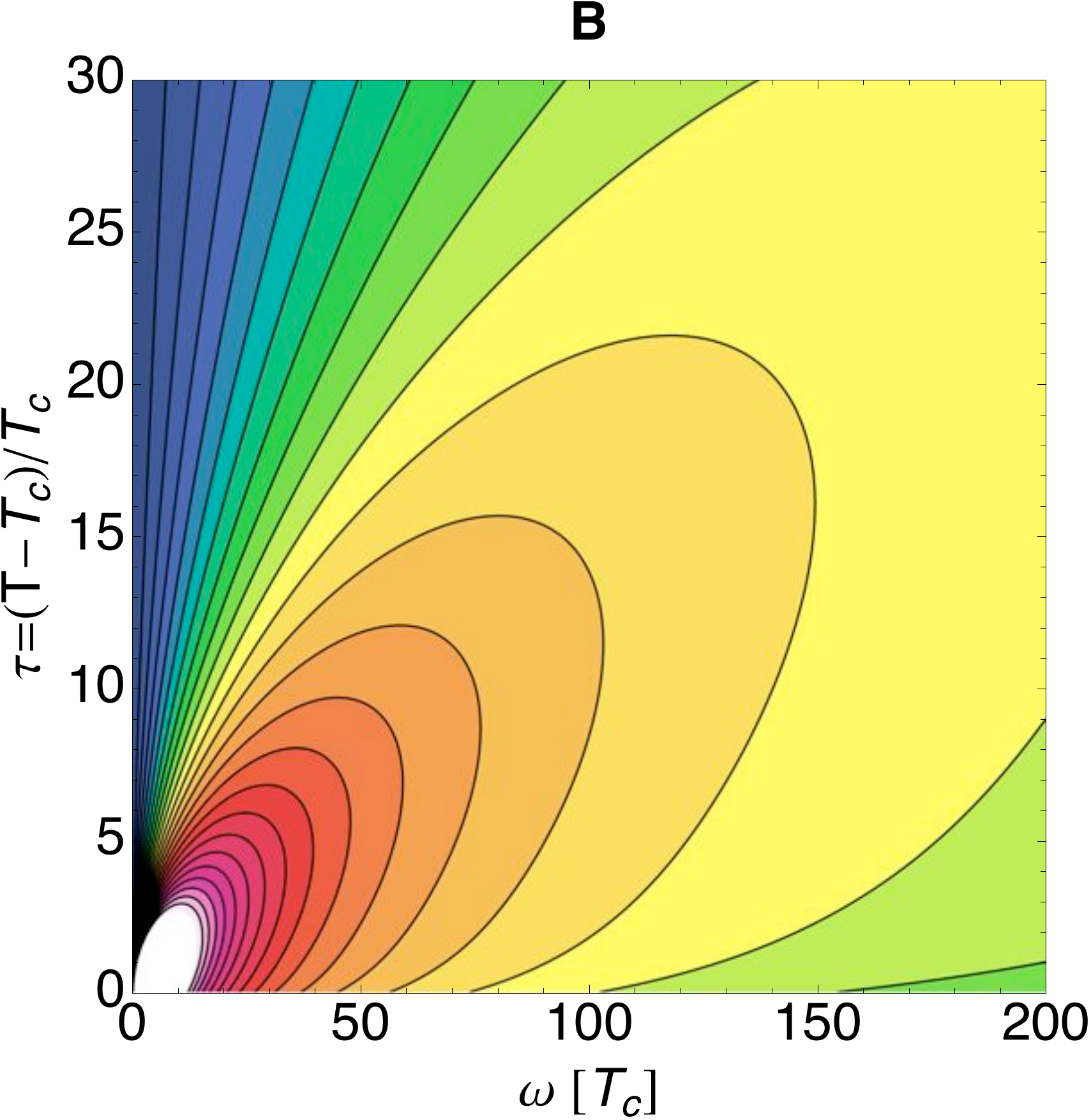}\\
\includegraphics[width=0.32\textwidth]{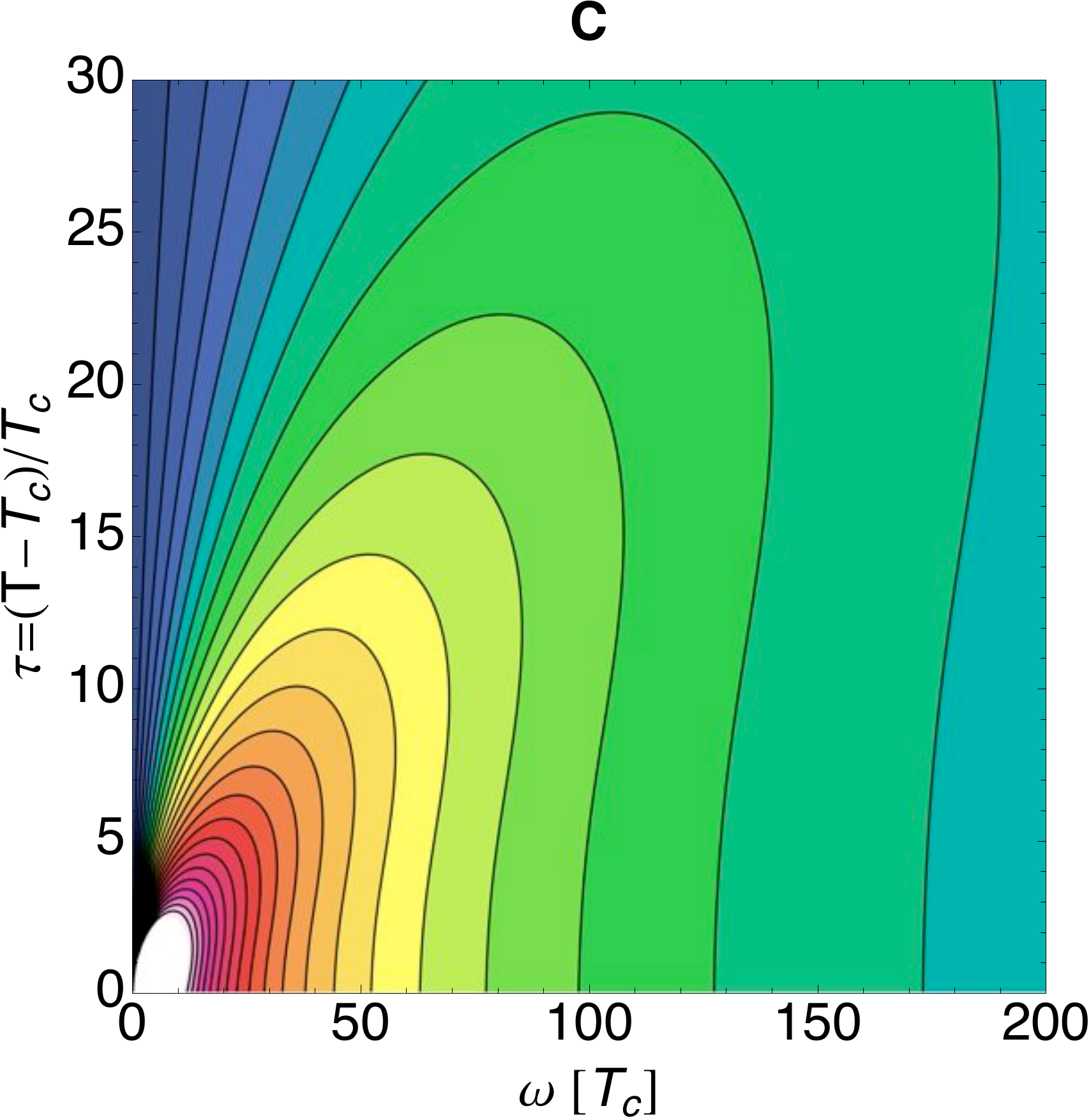}\quad
\includegraphics[width=0.32\textwidth]{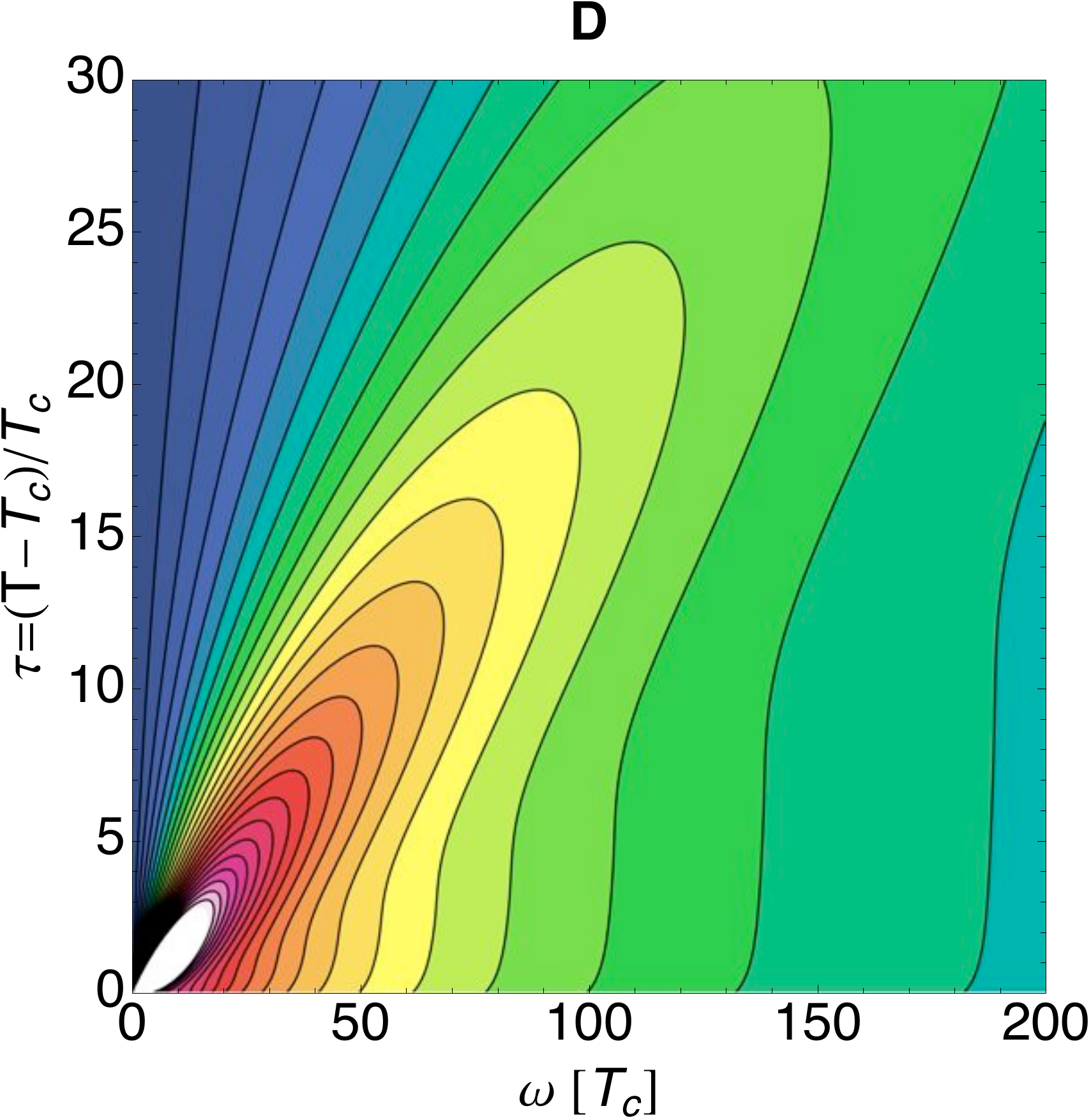}\quad
\includegraphics[width=0.32\textwidth]{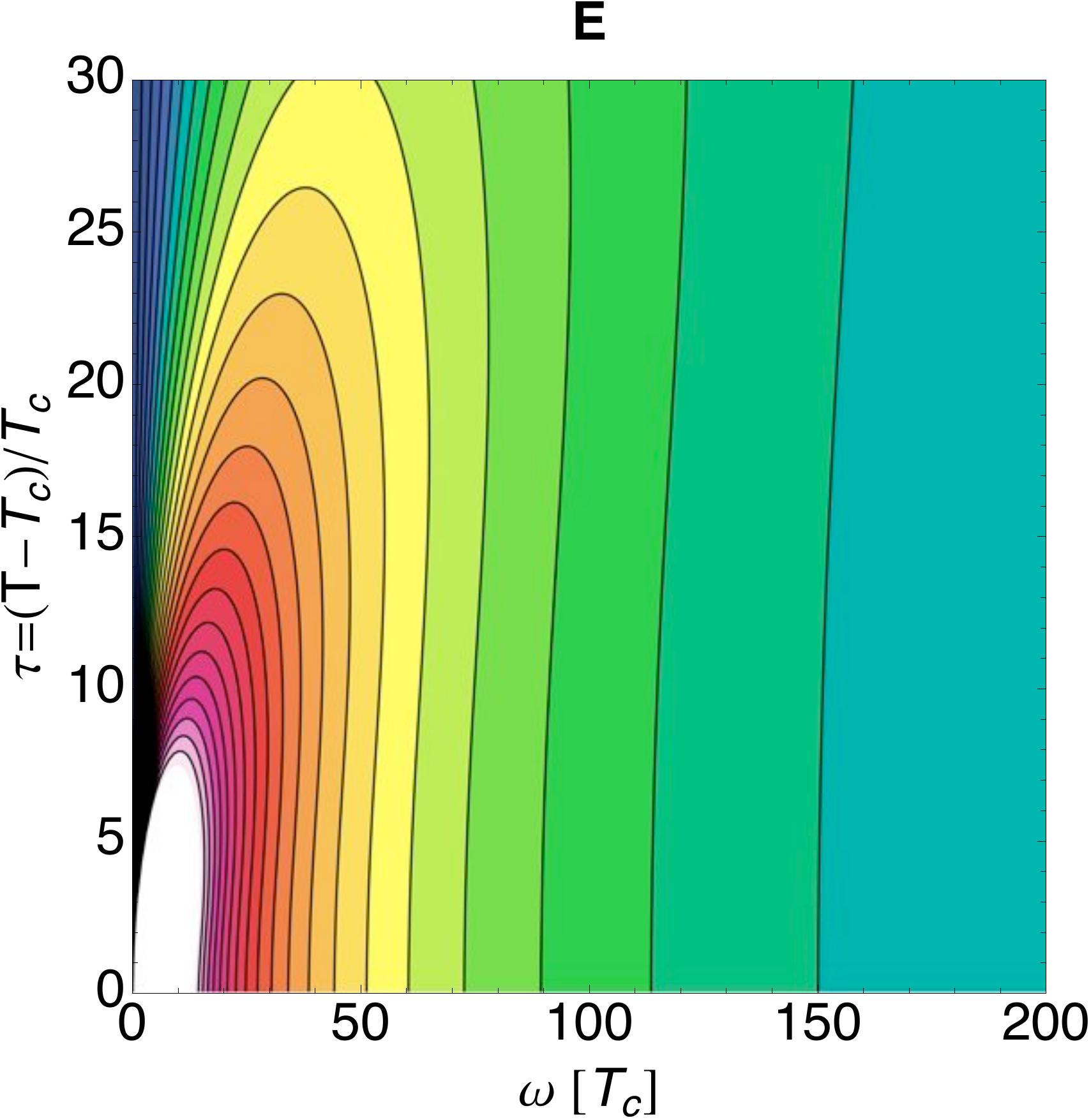}
\caption{(Color online) \textbf{Imaginary part of the pair susceptibility. A-E},
  False-color plot of the imaginary part of the pair susceptibility $\chi''(\omega,T)$ in arbitrary units as function of
  $\omega$ (in units of $T_c$) and reduced temperature $\tau=(T-T_c)/T_c$, for five different cases:  
case A represents the traditional Fermi liquid BCS theory (see section III case A with parameters
$T_c=0.01$, $g\approx0.39$, $\omega_b=0.45$%
), case B
is the Hertz-Millis type model with a critical glue (see section III case B with parameters
$\gamma=\frac{1}{3}$, $T_c=0.01$, $\Omega_0\approx0.0027$)%
,  case C is the phenomenological ``quantum critical BCS'' theory
(see section III case C with $\delta=\frac{1}{2}$, $T_c=0.01$, $g\approx0.19$, $\omega_b\approx0.1$, $x_0=2.665$)%
, case D corresponds to the ``large charge'' holographic 
superconductor with AdS$_4$ type scaling
(see section III case D with $\delta=\frac{1}{2}$, $T_c\approx0.40$, $e=5$)
  and case E is the ``small charge'' holographic 
superconductor with an emergent AdS$_2$ type scaling
(see section III case E with $\delta=\frac{1}{2}$, $T_c\approx 1.4\times 10^{-10}$, $e\approx 0$, 
$g=-\frac{17}{96}$, $\kappa\approx-0.36$)%
. $\chi''(\omega,T)$ should be directly proportional 
to the measured second order Josephson current (experiment discussed in the text). 
  In the bottom left of each plot is the
  relaxational peak that diverges (white colored regions are off-scale) as
  $T$ approaches $T_c$. This relaxational peak looks qualitatively
  quite similar for all five cases, while only at larger temperatures and
  frequencies qualitative differences between the five cases become manifest.
\label{fig:expsignal}}
\end{figure*}

Theoretically the pair susceptibility is defined as
\begin{align}
  \label{eq:1}
  \chi_{p}({\bf q},\omega)=-i\int_0^\infty dt e^{i\omega t -0^+ t}\langle\left[b^\dag({\bf q},0),b({\bf q},t)\right]\rangle,
\end{align}
where the Cooper pair order parameter $ b^\dag({\bf q},t)$ is built
out of the usual annihilation (creation) operators for electrons
$c^{(\dag)}_{{\bf k},\sigma}$ with momentum ${\bf k}$ and spin
$\sigma$. In the $s$-channel $ b^\dag({\bf q},t)=\sum_{\bf k}
c^\dag_{{\bf k}+{\bf q}/2,\uparrow}(t)c^\dag_{-{\bf k}+{\bf
    q}/2,\downarrow}(t)$. The imaginary (absorptive) part of this
susceptibility at zero-momentum is measured by the second order
Josephon effect. In figure \ref{fig:expsignal} we show the theoretical results for
standard BCS theory compared to four different limiting scenarios for
the quantum critical metallic state. This is our main result: the contrast is discernable by
the naked eye and this motivates our claim that this is an excellent
probe of the fundamental physics underlying the onset of
superconductivity. We 
will make clear that the specific temperature evolution of the dynamical pair susceptibility 
directly reflects the distinct RG flows underlying the superconducting
instability in each case.

\begin{figure*}
\includegraphics[width=2cm]{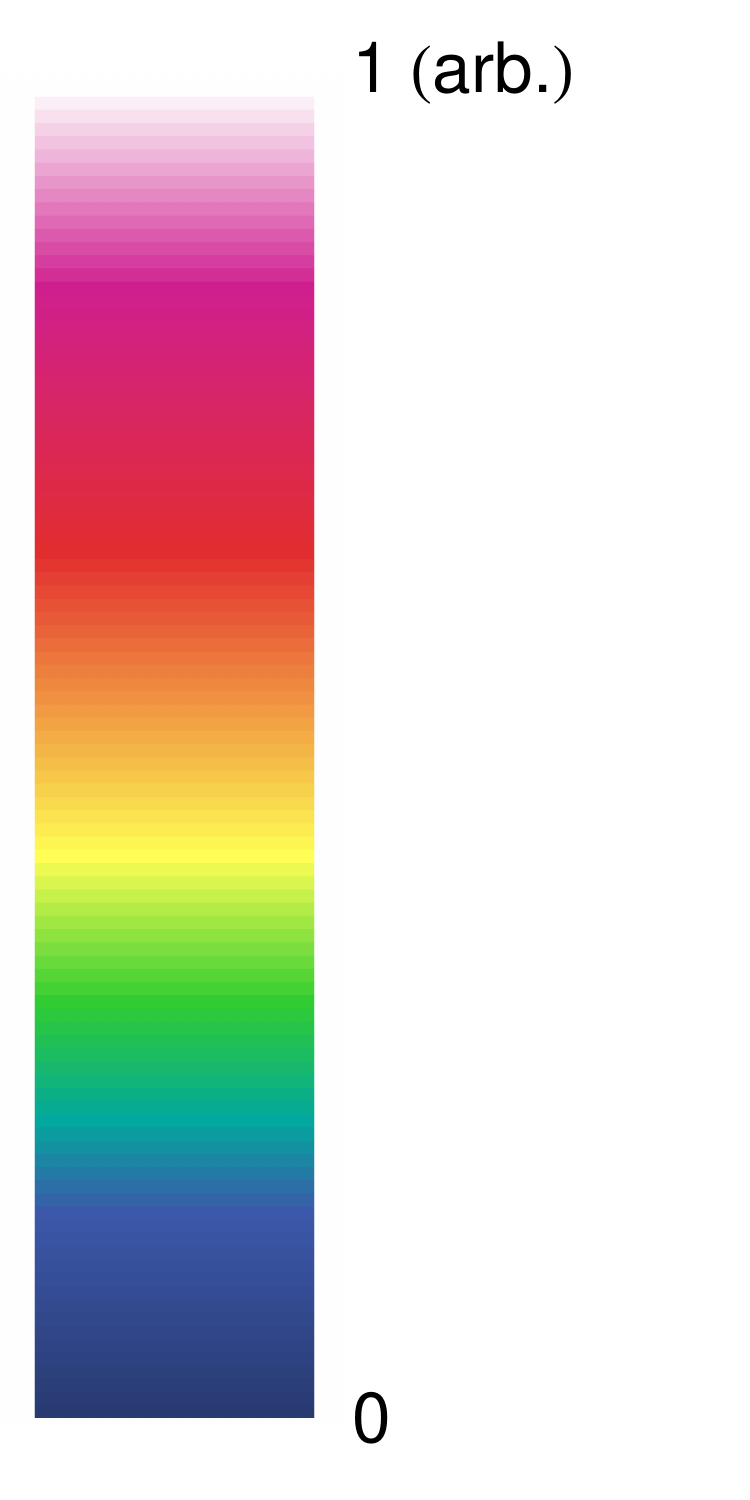}\quad
\includegraphics[width=0.32\textwidth]{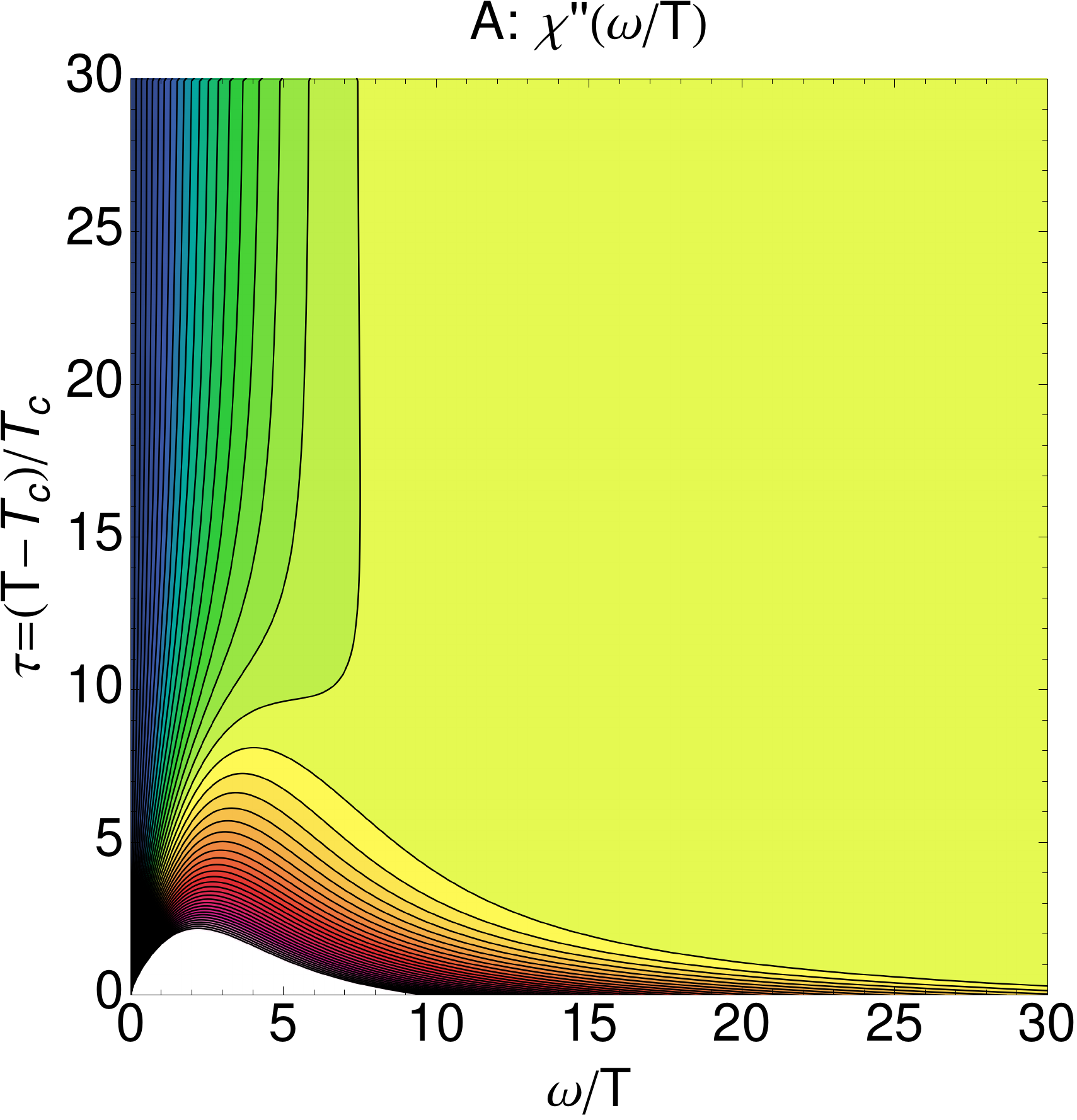}\quad
\includegraphics[width=0.32\textwidth]{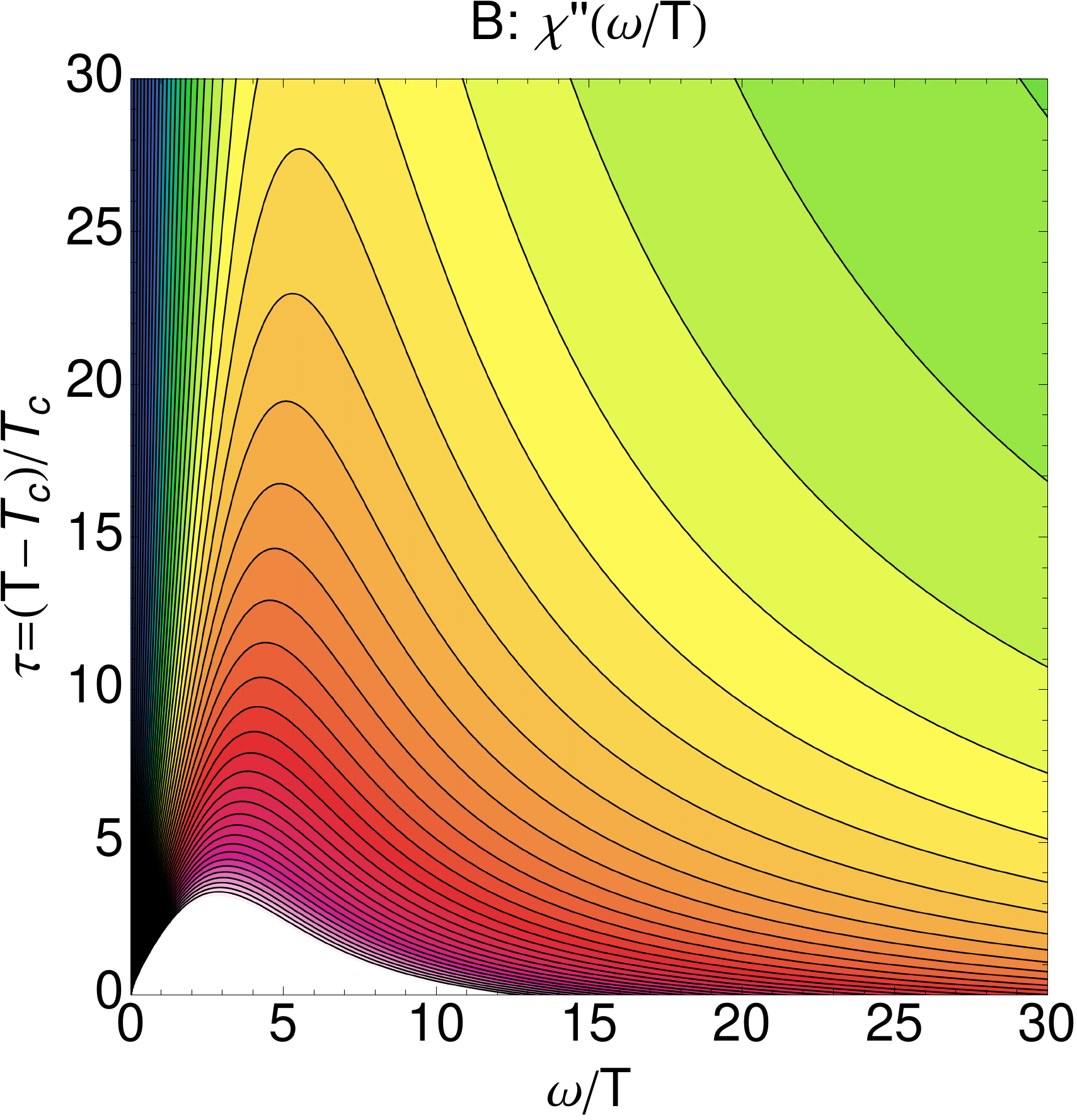}\\
\includegraphics[width=0.32\textwidth]{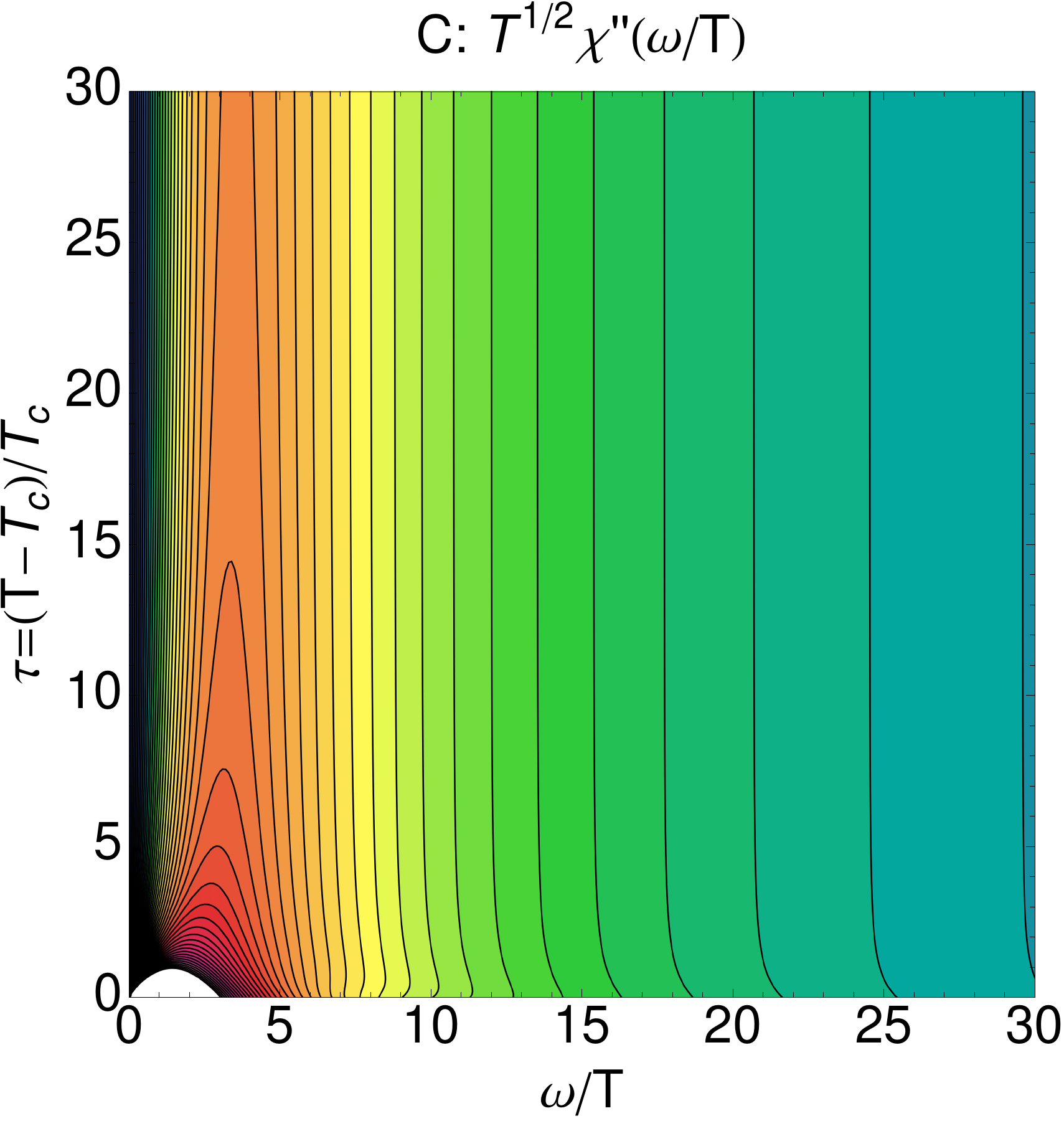}\quad
\includegraphics[width=0.32\textwidth]{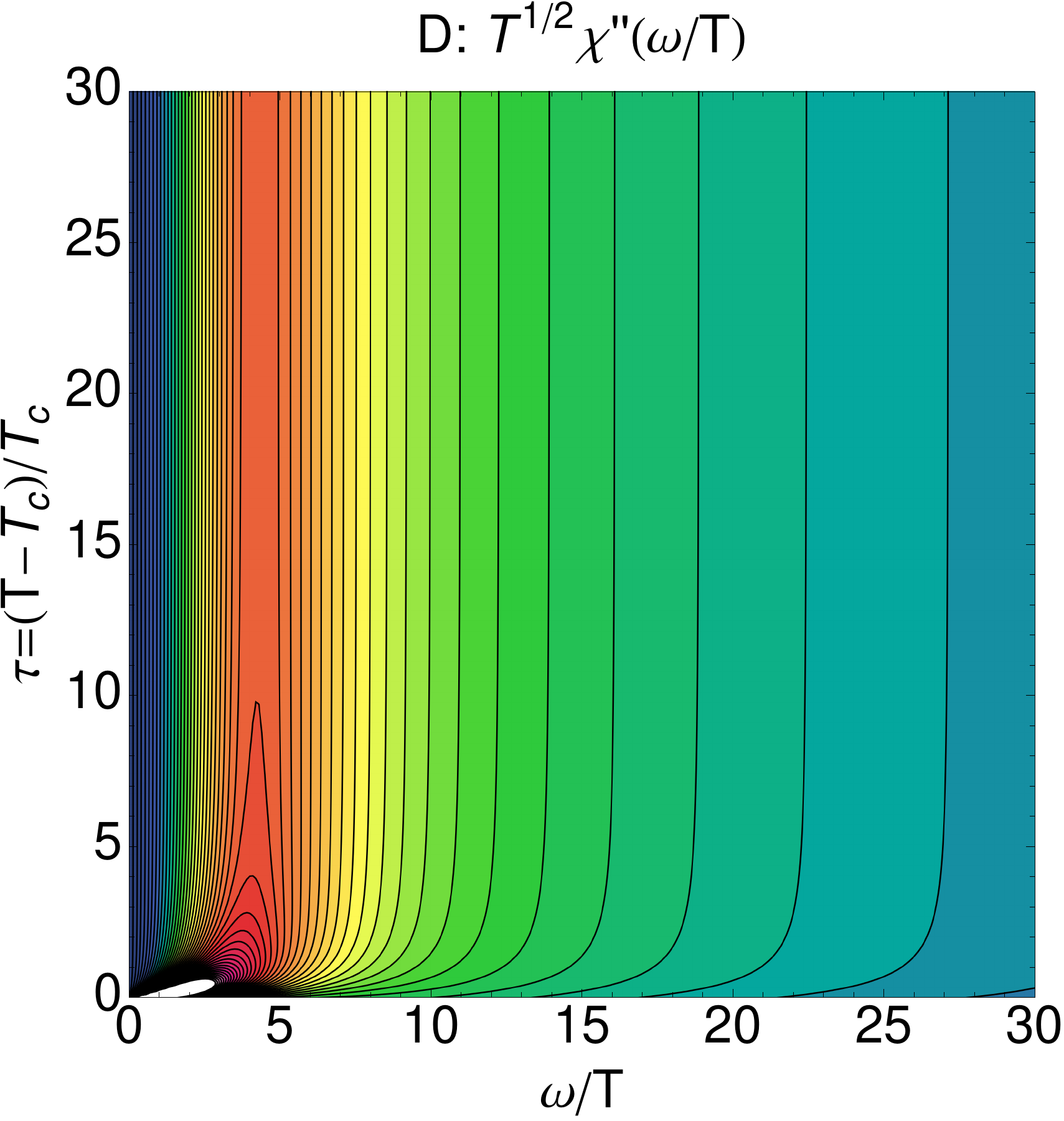}\quad
\includegraphics[width=0.32\textwidth]{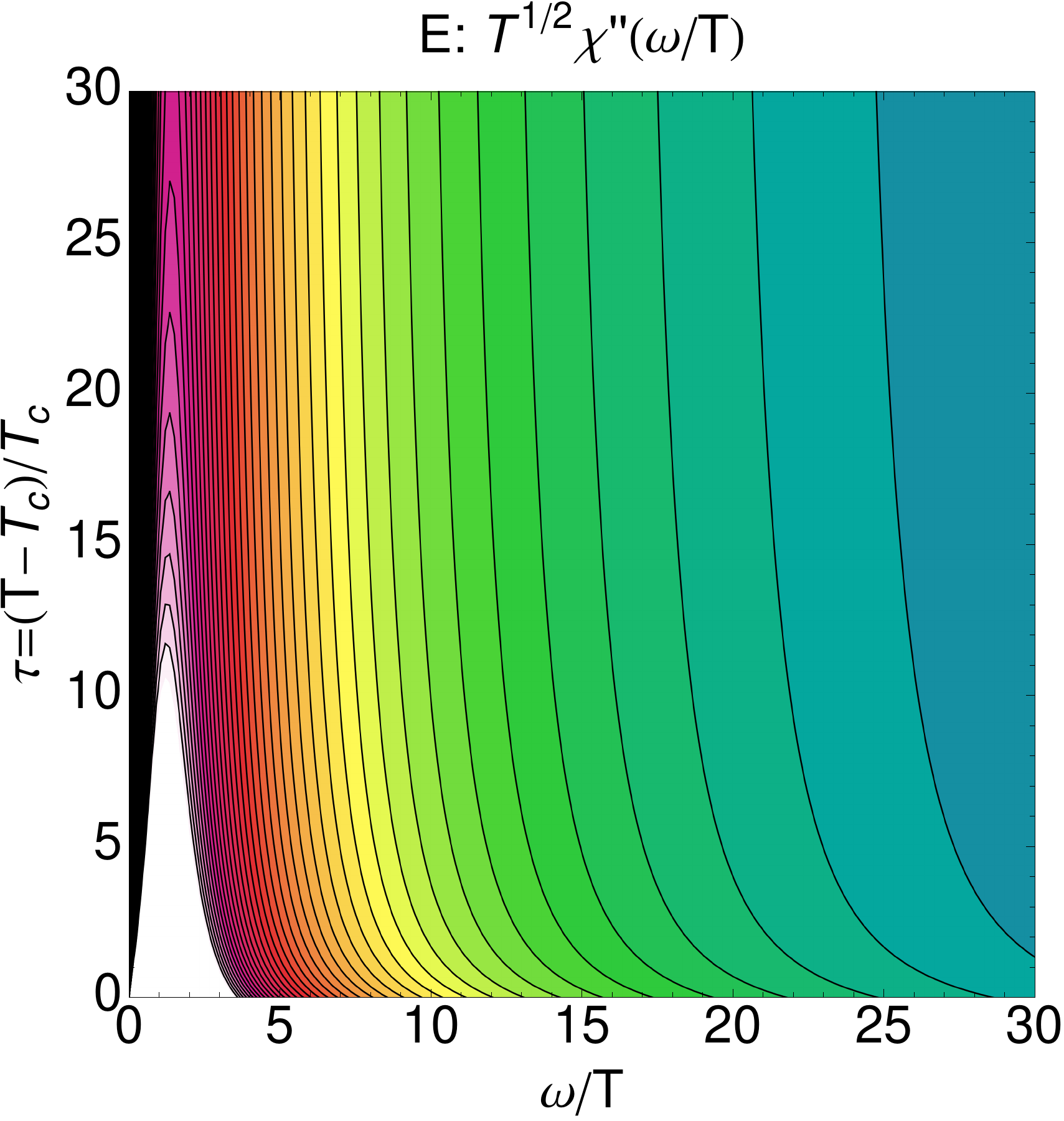}
\caption{(Color online) \textbf{Energy-temperature scaling of the pair
    susceptibility. A-E}, False-color plots of the imaginary part of
  the pair susceptibility, like in Fig.~\ref{fig:expsignal}, but now the horizontal axis is rescaled by temperature while the magnitude is 
rescaled by temperature to a certain power: we are plotting $T^\delta
  \chi''(\omega/T,\tau)$, in order to show energy-temperature scaling at high temperatures. 
  For quantum critical BCS (case C), AdS$_4$ (case D) and AdS$_2$ (case E), with a suitable choice of the exponent $\delta>0$, the contour lines run 
  vertically at high temperatures, meaning that the imaginary part of the pair susceptibility acquires a universal form 
  $\chi''(\omega, T)=T^{\delta}{\cal F}(\omega/T)$, with $\cal F$ a generic scaling function, the exact form of which depends on the choice of different models.
   Here we choose in cases C-D-E $\delta=1/2$, by construction. The weak coupling Fermi liquid BCS case A also shows scaling collapse at high temperatures, but with a marginal exponent $\Delta=0$. In the quantum critical glue model (case B) energy-temperature scaling
  fails: for any choice of $\delta$, at most a small fraction of the
  contour lines can be made vertical at high temperatures (here $\delta=0$ is displayed).
\label{fig:tempscaling}}
\end{figure*} 

In detail the five types (A-E) of pairing mechanisms whose susceptibility is given in Fig.~\ref {fig:expsignal} are:


Case A is based upon traditional Fermi liquid BCS theory and is included for comparison. The dynamical pair susceptibility is calculated through an Eliashberg-type  computation assuming a conventional Fermi-liquid interacting with  ``glue bosons''  in the form of a  single-frequency oscillator \cite{McMillan69,Scalapino69,Carbotte90}.
 Such a pair susceptibility would be found when the superconductivity would be due to ``superglue'' formed by bosons with a rather 
 well defined energy scale as envisaged in some 
 spin fluctuation scenarios \cite{Monthoux07,Scalapino11}.

Case B reflects the main stream thinking in condensed matter physics.
It
rests on the early work of Hertz \cite{Hertz76} and asserts that 
the essence of  BCS theory is still at work, i.e., 
 one can view the normal state at least in a perturbative sense as a Fermi-liquid, which coexists with a bosonic order parameter 
 field undergoing the quantum phase transition. The order parameter itself is Landau damped by the particle-hole
excitations, while the quantum critical fluctuations in turn couple strongly to the quasiparticles explaining the anomalous properties of the 
metallic state \cite{Lohneysen07}. Eventually the critical bosons cause the attractive interactions driving the pairing instability \cite{Chubukov04}. This notion is 
coincident with the idea that the pairing is due to spin fluctuations when the quantum phase transition involves magnetic order (as in
the heavy fermions and pnictides) while in the cuprate community  a debate rages at present concerning the role of other ``pseudogap'' 
orders like spontaneous currents and quantum nematics.
  The computation of the pair-susceptibility amounts to solving the full Eliashberg equations for a glue function that itself is algebraic in frequency $\lambda(\omega) \sim 1/ \omega^\gamma$ 
in the strong coupling regime as formulated by Chubukov and coworkers \cite{MoonChubukov10,Chubukov05,Chubukov03}. At first sight the 
resulting Fig.~\ref{fig:expsignal} B
looks similar to the remaining cases C-E that 
contain
more radical assumptions regarding 
the influence of the quantum scale invariance. However, as we will see, case B should leave a strong fingerprint in the data in the form of a strong violation 
of energy-temperature scaling (Fig.~\ref{fig:tempscaling}B).
 
Case C is a simple  phenomenological ``quantum critical BCS'' scaling theory \cite{She09}. 
 It is like BCS in the sense that a simple pairing glue 
is invoked but now it is assumed that the normal state is a non-Fermi
liquid which is controlled by conformal invariance. In other words, the `bare' pair propagator $\chi^0_\text{pair}(\omega,T)$,
 in the absence of glue, is described by a scaling
function.
The full pair susceptibility is then given 
by the RPA expression
\begin{equation}
\chi_\text{pair} ( \omega, T) = \frac{ \chi^0_\text{pair} (\omega, T)} { 1 - V \chi^0_\text{pair} (\omega, T)},
\label{RPA}
\end{equation}
where $V$ is the effective attractive interaction, that is non-retarded for simplicity. The pairing instability occurs when $ 1 - {V[\chi^0_\text{pair}(\omega=0,T_c)]'} = 0$. In quantum critical BCS one takes $\chi_\text{QBCS}^{0} (\omega) \sim 1/ (i
\omega)^{\delta}$, valid when $\omega\gg T$, as opposed to standard BCS where the bare fermion loop of the Fermi-gas yields a ``marginal''
pair propagator $\chi_\text{BCS}^{0} (\omega) = (1/E_F) [ \log ( \omega/E_F) + i ]$. 
One can now
deform the ``marginal'' Fermi liquid BCS case $\delta=0$ to 
``relevant'' pairing operators, i.e., with scaling exponent $\delta >0$. One effect of this power-law scaling is that $T_c$ becomes 
much larger. 
Our full calculations include finite temperature effects which serves as an IR cut-off and incorporate a retarded
nature of the interaction by considering an Eliashberg-style
generalization 
of equation~(\ref{RPA}).
Such power-law scaling behavior was recently identified in numerical dynamical 
cluster approximation calculations on the Hubbard model \cite{Jarrell10}. This was explained in terms of a marginal Fermi liquid (MFL),
i.e. the electron scattering rate proportional to the larger of temperature or frequency, in combination with a band structure
characterized by a van Hove singularity (vHS) which is precisely located at the Fermi energy \cite{Jarrell11}. 
The vHS is essential; a MFL self-energy added to standard BCS or crtical glue alone will not produce the power-law scaling.
The presence of a vHS can be measured
independently by ARPES \cite{Shen93, Campuzano94} and tunneling spectroscopy \cite{Fischer11} and therefore all the information is available in principle to distinguish this 
particular mechanism from the other cases. 
A careful study of the MFL pair susceptibility with both a smooth
density of states and vHS is included in Appendix B.

Cases D and E are 
radical departures of established approaches to superconductivity that emerged very recently from string theory. They are based on the anti-de-Sitter/conformal field theory correspondence (AdS/CFT) or ``holographic duality'' \cite{Maldacena97,Gubser98,Witten98}, asserting that
the physics of extremely strongly interacting quantum critical matter can be encoded in quasi-classical gravitational physics in a space-time with 
one more dimension. Including a charged black-hole in the center, a finite temperature and density is imposed in the field
theory, and the fermionic response of the resulting state is remarkably suggestive 
of
 the strange-metal behavior seen experimentally
in quantum critical metals. Although the (large-$N$ super-Yang-Mills) field theories
that AdS/CFT can explicitly address are remote to the physics of electrons in solids, there is much evidence suggesting that the correspondence describes generic ``scaling histories''. AdS/CFT can be viewed as a generalization of the Wilson-Fisher renormalization group that handles deeply-non-classical 
many-particle entanglements, for which the structure of the renormalization flow is captured in the strongly constrained gravitational physics of the 
holographic dual.  As such holography provides a new mechanism for superconductivity
: 
it requires, gravitationally encoded in black hole superradiance, that the finite density quantum critical metal turns into a superconducting 
state when temperature is lowered  \cite{Hartnoll08,Hartnoll0801}.  
This holographic
superconductivity (HS) is ``without glue'':  HS  is an automatism wired in the renormalization flow originating in the extreme thermodynamical 
instability of the uncondensed quantum critical metal at zero temperature. As we illustrate in Fig.~\ref{fig:expsignal}, AdS/CFT provides fundamentally new descriptions of the origin of superconductivity. The two cases D and E are the holographic analogues 
of local pair and ``BCS'' superconductors, in the sense that for the ``large charge'' case D the superconductivity sets in at a temperature of order of
the chemical potential $\mu$, while in the ``small charge''  case E the superconducting $T_c$ is tuned to a temperature that is small 
compared to $\mu$. 

The remainder of this paper is organized as follows. In section II, we
propose two explicit experimental approaches to measure the imaginary
part of the pairing susceptibility in the required temperature and
frequency range. One approach invokes modern thin film techniques and the other uses STM/STS/PCS techniques 
with a superconducting tip. Two heavy fermion systems, CeIrIn$_5$ and
$\beta$-YbAlB$_4$, are suggested as candidate quantum critical
superconductors. In section III, we present details of the
calculation of the pairing susceptibility in the five types of
models (A-E). For cases A-C, the full pair susceptibility is governed by the
Bethe-Salpeter equation, with the bare (electronic) pair susceptibility
and the pairing interaction (glue) as input. In the holographic
approaches D and E, the pair susceptibility is calculated from the
dynamics of the fluctuations of the dual scalar field in the AdS black hole
background in the dual gravity theory. The outcomes of these
calculations are further analyzed in section IV. Close to the
superconducting transition point, all the five models display
universal relaxational behavior. When moving away from T$_c$, one
detects sharp qualitative differences between the truly conformal
models (cases C-E) and the Hertz-Millis type models (case B). We
include in Section V our conclusions. There are two appendix
sections. In appendix A, the relaxational behavior of the holographic
models is derived using the near-far matching technique. In appendix
B, we present a Hertz-Millis type calculation of the pair
susceptibility in a marginal Fermi liquid.

\section{Proposed experimental setup}

In order to experimentally observe $\chi_{\rm pair}(\omega)$ via a second-order Josephson effect, one should measure the pair tunneling current, 
$I_{\rm pair}(V) \propto \chi_{\rm pair}''(\omega = 2eV/\hbar)$. This can be accomplished via a planar tunnel junction 
or weak link between the higher temperature superconductor ($T_c^{\rm high}$) and the probe superconductor ($T_c^\text{low}$). To extract the pair tunneling current from the total tunneling current the quasiparticle tunneling current contribution must be subtracted, e.g., by means of the Blonder-Tinkham-Klapwijk \cite{BTK82} formula and its ($d$-wave) generalizations. To minimize the masking effect of the quasiparticle current and to maximize the ranges of accessible reduced temperature and frequency the ratio $T_c^{\rm high}/T_c^{\rm low}$ of the two $T_c$'s should be as large as possible.


Perhaps the best candidate quantum critical superconductor is the heavy fermion system CeIrIn$_5$, since it appears to have a quantum
 critical normal state at ambient pressure, while its $T_c$ is a meager 0.4 K \cite{Kambe10}. The mixed valence compound $\beta$-YbAlB$_4$, which
  displays quantum criticality up to about 3 K without any tuning and becomes superconducting below 80 mK \cite{Matsumoto11}, is another possible choice. 
  The challenge is now to find a good insulating barrier that in turn is well connected to a ``high'' $T_c$ source superconductor. One option for the latter is the $T_c=\text{40 K}$ MgB$_2$ system; 
an added difficulty is that one should take care that this $s$-wave superconductor can form a Josephson contact with the non-conventional (presumably $d$-wave) quantum critical superconductor.
This has on the other hand the great advantage that the quasiparticle current is largely suppressed because of the 
presence of the full gap, compared to an unconventional source superconductor with its nodal quasiparticles. As a start, one could employ the modern material fabrication techniques 
of monolithic molecular beam epitaxy (MBE) \cite{Shishido10} and pulsed laser deposition (PLD) \cite{Bergeal08}, to form a junction between MgB$_2$ 
and Al with an insulating  aluminum-oxide junction layer. Reduced temperatures $\tau=(T -T_c)/T_c$ 
up to 40 with low noise $\omega$-values into the mV regime could be obtained with these two $s$-wave superconductors. 

A more challenging technique is to utilize the recent advances in scanning tunneling microscopy and spectroscopy (STM/STS) \cite{Fischer07}
 and point contact spectroscopy (PCS) \cite{Greene08} to form or glue a tiny crystal or whisker of YBa$_2$Cu$_3$O$_{7-y}$ ($T_c^{\rm high} =\text{90 K}$)  to 
 a normal Ir or Pt tip and tunnel or weakly contact the tip to the heavy fermion superconductor through 
 its freshly cleaved surface. With the enormous spread in transition temperatures $\tau$-values of over 100 could be reached
  within a mV low-noise region for two such $d$-wave superconductors. 
  
There are certainly difficulties with the cuprate superconductors such as surface charging, gap-reduction and low Josephson currents.
 These troublesome issues could be resolved by using a pnictide superconductor tip \cite{Noat10} or a combination of a hole-doped HTS ($T_c^{\rm high}$) 
 and concentration-tuned Nd$_{2-x}$Ce$_x$CuO$_{4-\delta}$ ($T_c^\text{low}<24$ K), an electron-doped superconductor, to increase the Josephson current.
  Stimulated by our pair-susceptibility calculations, we trust the challenged experimentalists will evaluate the above possibilities 
  in their efforts towards novel thin film and tunneling spectroscopy investigations.

\section{Calculating the pair susceptibility for different models}

\subsection*{Cases A-C: Pairing mechanisms with electron-glue dualism}

The pair susceptibility is a true two-particle quantity, i.e., it is derived from the full two-particle (four point) Green's function which is traced over external fermion legs: let $\chi(k,k';q)$ be the full four-point correlation function with incoming momenta/frequencies $(-k, k+q)$ and outgoing momenta/frequencies $(-k',k'+q)$, then the pair susceptibility $\chi_\text{pair}(i\Omega,{\bf q})=\sum_{k,k'}\chi(k,k';q)$. Here momentum and frequency are grouped in a single symbol $k=({\mathbf k},i\omega)$ and we formulate equations using Matsubara frequencies.
 
The full pair susceptibility includes contributions from all forms of interactions. One commonly used approximation strategy is to separate it into two parts: an electronic part and a glue part. The glue is generally considered to be retarded in the sense that it has a characteristic energy scale $\omega_b$ that is small compared to the ultraviolet cut-off scale $\omega_c$. Under this retardation assumption, i.e., a small Migdal parameter, the electron-glue vertex corrections can thus be ignored and the effects of the glue can be described by a Bethe-Salpeter-like equation in terms of the `vertex' operators $\Gamma(k;q)=\sum_{k'}\chi(k,k';q)$, i.e., a partial trace over $\chi(k,k';q)$.
Further simplification can be made by assuming that the pairing problem in quantum critical metals can still be treated within the Eliashberg-type theory, with the electronic vertex operator $\Gamma_0$ and the glue propagator $D$ strongly frequency dependent, but without substantial momentum dependence. The glue part will only appear in the form of a frequency-dependent pairing interaction $\lambda(i\Omega)=\int d^d{\mathbf q}D({\mathbf q};i\Omega)$.
The Bethe-Salpeter equation (or Dyson equation for the four point function) then reads
\begin{equation}
\Gamma(i\nu;i\Omega)=
\Gamma_0(i\nu;i\Omega)+{\cal A}\,\Gamma_0(i\nu;i\Omega)\sum_{\nu'}\lambda(i\nu'-i\nu)\Gamma(i\nu';i\Omega),
\label{Master}
\end{equation}
at $ {\mathbf q}=0$. Note that the pair susceptibility is a bosonic response, hence $i\Omega$ is a bosonic Matsubara frequency whereas $i\nu$ is fermionic. For given electronic part $\Gamma_0(i\nu;i\Omega)$ and glue part $\lambda(i\Omega)$ equation (\ref{Master}) can be solved, either by iteration or by direct matrix inversion.
A further frequency summation over $\nu$ of $\Gamma$ finally yields the full pair susceptibility $\chi_{\rm pair}(i\Omega,{\bf q}=0)=\sum_{\nu}\Gamma(i\nu;i\Omega)$ at imaginary frequency $i\Omega$. The superconducting transition happens when the real part of the full pair susceptibility at $\Omega=0$ diverges. 
To obtain the desired real-frequency dynamical pair susceptibility, a crucial step is the analytic continuation, i.e., the replacement $i\Omega\to\omega+i 0^+$. We choose the method of analytic continuation through Pad\'e approximants via matrix inversion \cite{baker80,vidberg77,beach00}, which performs remarkably well in our case, likely due to the fact that here the pair susceptibility is a very smooth function with only a single characteristic peak/feature.

Different models are characterized by different $\Gamma_0(i\nu;i\Omega)$ and $\lambda(i\Omega)$. 
We will present the three non-holographic approaches to pairing, i.e., cases A--C,  in the remainder of this section.

\subsubsection{Case A: Fermi liquid BCS}
We consider a free Fermi gas, interacting via a normal glue, say an Einstein phonon, for which the pairing interaction is of the form
\begin{equation}
\lambda(i \Omega)=\frac{g}{\cal A}\frac{\omega_b^2}{\omega_b^2+\Omega^2}.
\label{one-phonon}
\end{equation} 
 For the Fermi gas  Wick's theorem applies, and the electronic part of the pair susceptibility is simply the convolution of single-particle Green's functions,
\begin{equation}
\chi_{\text{pair},0}({\mathbf q},i\Omega)=\frac{T}{N}\sum_{{\mathbf k},n}G\left(-{\mathbf k},-i\nu_n\right)G\left({\mathbf k}+{\mathbf q},i\nu_n+i\Omega\right).
\label{fermionloop}
\end{equation}
If we ignore self-energy corrections we may substitute the free fermion Green's function $G({\mathbf k},i\omega)=1/(i\omega_n-\varepsilon_{\mathbf k})$. The imaginary part of the bare pair susceptibility then has the simple form $\chi_0''(\omega)=\frac{1}{\omega_c}\tanh\!\left(\frac{\omega}{4T}\right) $ at $\mathbf q=0$. Here the Fermi energy acts as the ultraviolet cut-off, with $\omega_c=\frac{2}{\pi N(0)}\simeq E_F$. The electronic vertex operator reads
\begin{equation}
\Gamma_0(i\nu_n, i\Omega)=\frac{2 T}{\omega_c(2\nu_n+\Omega)} \left[ \theta(\nu_n+\Omega)-\theta(-\nu_n) \right]=\frac{2 T}{\omega_c}\left|\frac{\theta(\nu_n+\Omega)-\theta(-\nu_n) }{2\nu_n+\Omega}\right|,
\label{gamma0-bcs}
\end{equation}
with $\theta(x)$ the Heaviside step function.

A full Eliashberg treatment includes self-energy corrections and modifies equation~(\ref{gamma0-bcs}) to
\begin{equation}
 \Gamma_0(i\nu_n, i\Omega)=\frac{2 T}{\omega_c} \left|\frac{\theta(\nu_n+\Omega)-\theta(-\nu_n)}{ (\nu_n+\Omega)Z(\nu_n+\Omega)+\nu_nZ(-\nu_n)}\right|,
\label{gamma0-cg}
\end{equation}
where $\omega_n Z(\omega_n)\equiv\omega_n+\Sigma(i\omega_n)$. For small and non-singular pairing interaction $\lambda(i\Omega)$ the effect of the self-energy corrections will be minor.

\subsubsection{Case B: Critical Glue BCS}
In this subsection, we replicate one class of scenarios which attribute the novelty of unconventional superconductivity in such systems to the peculiar behavior of the glue when approaching the QCP. The glue part is assumed to become critical near the QCP, while the electronic part is kept a fermion bubble as in conventional BCS theory, equation~(\ref{fermionloop}), with self-energy corrections included. This class of scenarios are arguably best represented by the models introduced by Chubukov and collaborators \cite{MoonChubukov10}, where they assume that pairing is mediated by a gapless boson, and the pairing interaction is of the power-law form
\begin{equation}
\lambda(i\Omega)=\Bigl(\frac{\Omega_0}{|\Omega|}\Bigr)^\gamma.
\label{critical-pairing}
\end{equation}
Here the exponent $0<\gamma<1$ parameterizes the different models. The pairing interaction has a singular frequency dependence, which makes the pairing problem in such models qualitatively different from that of the Fermi liquid BCS model. The coupling strength is absorbed in the parameter $\Omega_0$, which is the only scale-full parameter in this model. Thus the superconducting transition temperature should be proportional to $\Omega_0$, with a model-dependent coefficient, $T_c=A(\gamma)\Omega_0$.

The massless boson contributes a self-energy $\Sigma(i\omega_n)$ to the electron propagator,
\begin{equation}
 \Sigma(i\omega_n)=\omega_n\Bigl(\frac{\Omega_0}{|\Omega|}\Bigr)^\gamma S(\gamma,n),
 \end{equation}
where $S(\gamma,n)=|n+1/2|^{\gamma-1}\left[ \zeta(\gamma)-\zeta(\gamma,|n+\frac{1}{2}|+\frac{1}{2}) \right]$, with $\zeta(\gamma)$ the Riemann zeta function and $\zeta(\gamma,n)$ the generalized  Riemann zeta function. 


The presence of the scale-full parameter $\Omega_0$ will generically prevent simple energy-temperature scaling of $\chi_\text{pair}(\omega,T)$. Only for the limits $T\ll \Omega_0$ or $T\gg \Omega_0$ one should recover energy-temperature scaling.

\subsubsection{Case C: Quantum Critical BCS}
In this subsection we will consider the scenario of quantum critical BCS \cite{She09}, where the novelty of unconventional superconductivity is attributed solely to the peculiar behavior of the electronic part in the quantum critical region, with the glue part assumed featureless. For the glue part we will use, as in the Fermi liquid BCS case, the smooth and nonsingular pairing `Einstein phonon' interaction, equation~(\ref{one-phonon}), to calculate the dynamical pair susceptibility in the QCBCS scenario.
The quantum criticality is entirely attributed to the electronic part, i.e., the `bare' pair susceptibility is assumed to be a conformally invariant state and is considered to be a relevant operator in the renormalization flow sense. In other words, this amounts to the zero temperature power-law form $\chi_{\text{pair},0}''(\omega,T=0)={\cal A} \omega^{-\alphap}$, with $0<\alphap<1$. At finite temperature, the electronic part of the pair susceptibility can be expressed as a scaling function,
\begin{equation}
\chi_{\text{pair},0}(\omega,T)=\frac{Z}{T^{\alphap}}{\cal F}\left( \frac{\omega}{T} \right),
\end{equation}
which, in the hydrodynamical regime ($\hbar\omega\ll k_BT$) reduces to
$\chi_{\text{pair},0}(\omega,T)=\frac{Z'}{T^{\alphap}}\frac{1}{1-i\omega\tau_{\rm rel}}$,
 with $\tau_{\rm rel}\approx\hbar/k_BT$. 
Note that the Fermi liquid is the corresponding marginal case $\delta=0$ with $\chi_{\text{pair},0}''(\omega, T=0)\simeq{\rm constant}$. With a relevant scaling exponent $\alphap$ on the other hand, more spectral weight is accumulated at lower energy scales, where pairing is more effective. The gap equation becomes algebraic instead of exponential, and this implies that even a weak glue can give rise to a high transition temperature. 


The QCBCS scenario is a phenomenological theory; in the absence of a microscopic derivation of the scaling function $\mathcal{F}(\omega/T)$ a typical functional form is chosen. One example of such a typical scaling function ${\cal F}(\omega/T)$ that possesses the above two limiting forms at low and high temperatures can be found in $1+1$-dimensional conformal field theories, ${\cal F}''(y)=\sinh(\frac{y}{2})B^2(s+i\frac{y}{4\pi},s-i\frac{y}{4\pi})$,
where $B$ is the Euler beta function, and $s=1/2-\delta/4$. Another example, which will be used to calculate the full pair susceptibility in this paper, is a simple generalization of the free fermion vertex operator, equation~(\ref{gamma0-bcs}),
\begin{equation}
\Gamma_0(i\nu_n, i\Omega)= \frac{(1-\alpha)T}{\omega_c^{1-\alpha}} \frac{|\theta(\nu_n+\Omega)-\theta(-\nu_n)|}{|2\nu_n+\Omega|^{\alpha+1}},
\label{gamma0-qc}
\end{equation}
\begin{equation}
\chi_{\text{pair},0}(i\Omega)=\sum_{\nu_n}\Gamma_0(i\nu_n, i\Omega)=\frac{(1-\alpha)T}{\omega_c^{1-\alpha}}\frac{2}{(4\pi T)^{1+\alpha}}\;\zeta\left(1+\alpha,\frac{1}{2}-i \frac{i\Omega}{4\pi T}\right)
.
\end{equation} 
Here $\zeta$ is again the generalized (Hurwitz) zeta function. Since analytic continuation is trivial it is easy to confirm 
 that this choice of vertex operator produces a  relevant bare pair susceptibility with $\alphap=\alpha$, a power-law tail at high frequency, and the linear hydrodynamic behavior at low frequency. There is a single peak at frequencies of order the temperature, the precise location of which we may fine-tune by introducing a parameter $x_0$ (defined as the argument of the scaling function $\mathcal{F}(x)$ at which the low-frequency linear and high frequency power-law asymptotes would cross). 

We would like to emphasize again that QCBCS is a phenomenological theory: equation~(\ref{gamma0-qc}) is an educated guess for what a true conformally invariant two-particle correlation function (partially traced) may look like. However, combined with a glue function, equation~(\ref{one-phonon}), it is perfectly valid input for the Eliashberg framework, i.e., the Bethe-Salpeter equation~(\ref{Master}), and  delivers quite a hight $T_c$.

\subsection*{Cases D-E: Holographic superconductivity}

In the holographic approach to superconductivity  the 2+1 dimensional conformal field theory (CFT) describing the physics at the quantum critical point is encoded in a 3+1 dimensional string theory in a spacetime with a negative cosmological constant (anti-de Sitter space) \cite{Maldacena97,Witten98,Gubser98}. In a ``large $N$, strong coupling limit'' this string theory can be approximated by classical general relativity in an asymptotically anti-de Sitter (AdS) background coupled to various other fields.
Most importantly, a precise dictionary exists how to translate properties of the AdS gravity theory to properties of the CFT including the partition function. In particular, a global symmetry in the CFT is a local symmetry in the gravity theory with the boundary-value of the gauge field identified with the source for the current in the CFT. This provides the set-up for holographic superconductivity in the standard approximation where superconductivity is studied as the spontaneous symmetry breaking of a global $U(1)$, that is subsequently weakly gauged to dynamical electromagnetism.

\subsubsection{Case D: ``Large charge'' AdS$_4$ holographic superconductor}

The simplest model to obtain a holographic superconductor is therefore
Einstein gravity minimally coupled to a $U(1)$ Maxwell
field $A_{\mu}$ and a charged complex scalar $\Psi$ with charge $e$
and mass $m$.\cite{Gubser08, Hartnoll08, Hartnoll0801} The charged scalar will be dual to the order parameter in the CFT --- the pairing operator. Since the underlying field theory is strongly coupled there is no sense in trying to identify the order parameter as some ``weakly bound'' pair of fermions and we ought to study the order parameter directly.

This system has the action
\be \mathcal{S}_0=\int d^4x \sqrt{-g}\bigg[R+\frac{6}{L^2}-\frac{1}{4}F_{\mu\nu}F^{\mu\nu}-m^2|\Psi|^2-
|\nabla^\mu\Psi-ieA^\mu\Psi|^2\bigg],
\label{action0}\ee
where $R$ is the Ricci scalar and the AdS radius $L$ can be set to $1$.  The charged AdS Reissner-Nordstr\"{o}m (RN) black hole is a solution with $\Psi=0$. This solution has the spacetime metric and electrostatic potential 
 \bea\label{rnbh}
ds^2&=&-f(r)dt^2+\frac{dr^2}{f(r)}+r^2(dx^2+dy^2),\nonumber\\
f(r)&=&r^2-\frac{1}{r}\bigg(r_+^3+\frac{\rho^2}{4r_+}\bigg)+\frac{\rho^2}{4r^2},\\
A&=&\rho\bigg(\frac{1}{r_+}-\frac{1}{r}\bigg)dt,\nonumber\eea
where $r_+$ is the position of the horizon and $\rho$ corresponds to the charge density of the dual field theory. 
The temperature of the dual field theory is identified as the Hawking temperature of the black hole $T=\frac{3r_+}{4\pi}(1-\frac{\rho^2}{12r_+^4})$, and the chemical potential is $\mu=\rho/r_+$.  The AdS-RN solution preserves the $U(1)$ gauge symmetry and corresponds holographically to the CFT in a state at finite temperature and chemical potential. 

The essence of holographic superconductivity is that 
below some critical temperature $T_c$, the charged AdS-RN black hole 
becomes unstable and develops a non-trivial (normalizable) scalar condensate, i.e., $\Psi\neq0$, which breaks the $U(1)$ gauge symmetry. The asymptotic $r \rightarrow \infty$ value of $\Psi$ is the value of the order parameter in the  CFT. Thus in the dual field theory  a global U(1) symmetry is broken correspondingly.  Such a minimal model therefore naturally realizes ($s$-wave) superconductivity \cite{Hartnoll08,Hartnoll0801}.

Using explicit details of the AdS/CFT dictionary, the dynamical susceptibility of the spin-zero charge-two order parameter ${\mathcal{O}}$ in the boundary field theory can be calculated from the dynamics of the fluctuations of the corresponding scalar field $\Psi$ in the AdS black hole background in the gravity side. At zero momentum, we can expand $\delta\Psi$ as $\delta\Psi(r,x,y,t)|_{k=0}=\psi(r)e^{-i\omega t}$.
The equation of motion for $\psi(r)$ is
\be\label{eompsi} \psi''+\bigg(\frac{f'}{f}+\frac{2}{r}\bigg)\psi'+\bigg(\frac{(\omega +eA_t)^2}{f^2}-\frac{m^2}{f}\bigg)\psi=0.\ee
We are interested in the retarded Green's function. This translates into imposing infalling boundary condition at the horizon \cite{Son02}, i.e.,
$\psi(r) \simeq (r-r_+)^{-i\frac{\omega}{4\pi T}}$, as $ r\to r_+$. The CFT Green's function is then read off from the behavior of solutions $\psi_\mathrm{sol}$ to (\ref{eompsi}) at spatial infinity $r\rightarrow \infty$. Near this AdS boundary, one has
$\psi(r) \simeq \frac{\psi_-}{r^{\vartriangle_-}}+\frac{\psi_+}{r^{\vartriangle_+}}$,
where $\vartriangle_\pm=\frac{3}{2}\pm \nu$ with $\nu=\frac{1}{2}\sqrt{9+4m^2}.$ We focus on the case $0<\nu<1$, where both modes ${\psi_\pm}$ are normalizable. We furthermore choose ``alternate quantization'' with ${\psi_+}$ as the source and ${\psi_-}$ as the response, such that in the large frequency limit the order parameter susceptibility behaves as $1/\omega^{2\nu}$. In that case, the Green's function is given by
 \cite{Klebanov99,Son02}
\be \chi_{\rm pair}=\mathcal{G}^R_{\mathcal{O}_{-}^\dag\mathcal{O}_{-}}\sim - \frac{\psi_-}{\psi_+}.\ee 

\begin{figure}[ht]
\begin{center}
\includegraphics[width=8cm, clip]{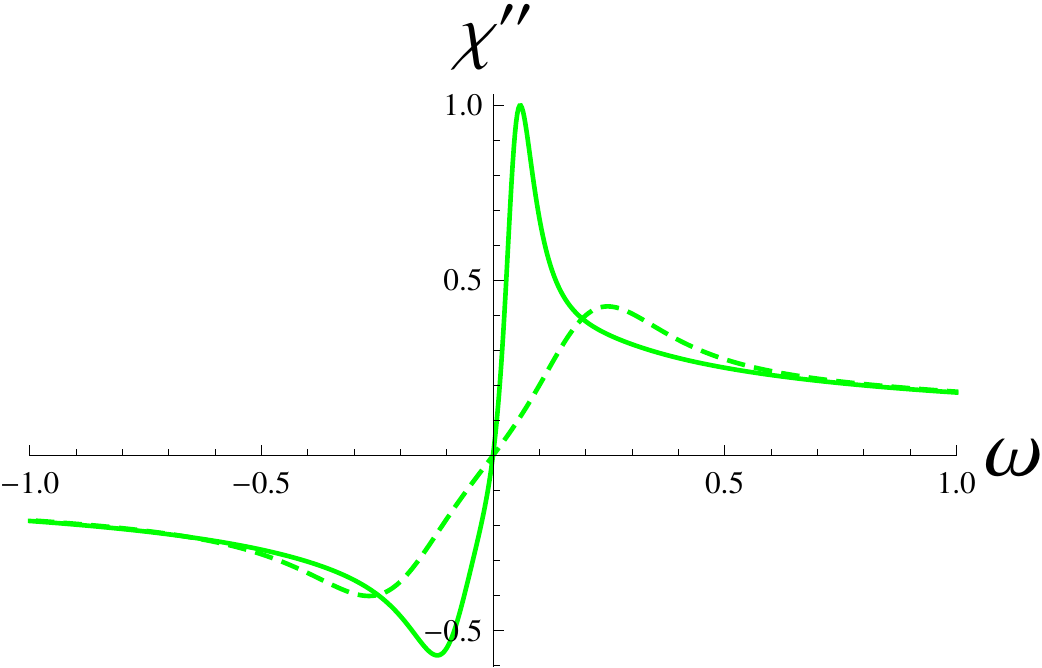}
\includegraphics[width=8cm, clip]{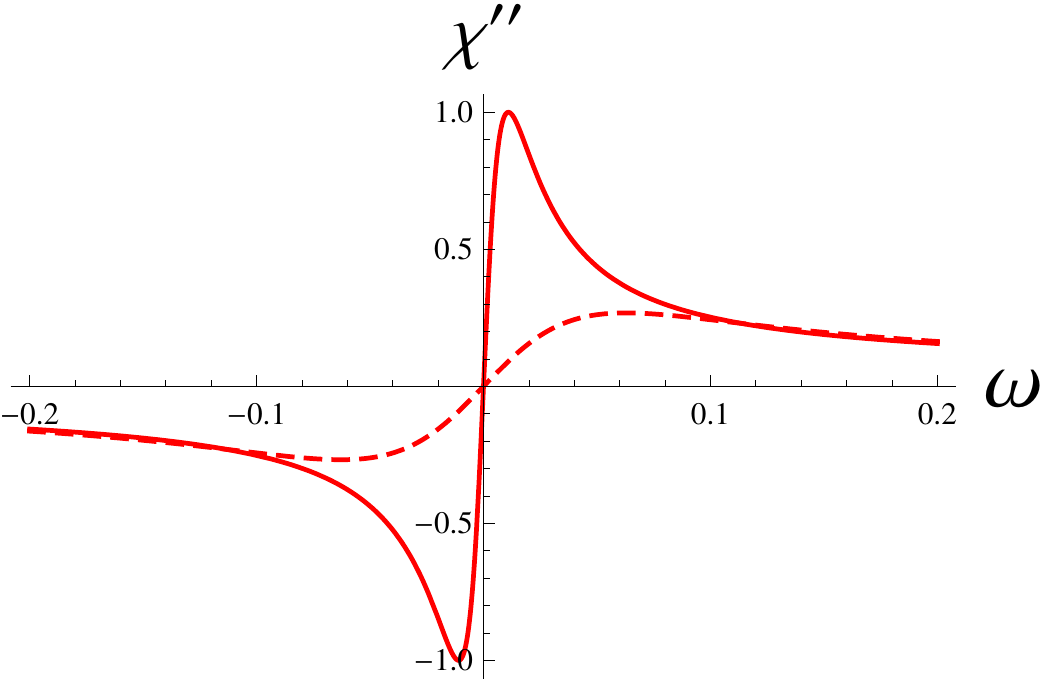}
\end{center}
\caption{(Color online) {\bf Particle-hole (a)symmetry of the relaxational peak.} Particle-hole (a)symmetry as seen from the line shape of $\chi''(\omega)$ for the two different kinds of holographic superconductors: local pair AdS$_4$ (left) and BCS-type AdS$_2$ (right). The solid lines correspond to reduced temperature $\tau=(T-T_c)/T_c=1$ and the dashed lines correspond to $\tau=5$. The AdS$_4$ case has a particle-hole asymmetric pair susceptibility, while this symmetry is restored in the AdS$_2$ case. 
\label{asymmetry}}
\end{figure}

From equation (\ref{eompsi}), the boundary conditions at the horizon and the dictionary entry for the Green's function, the order parameter susceptibility has the manifest symmetry $\chi(\omega,e)=\chi^*(-\omega,-e)$. This implies generic particle-hole asymmetry as for $e\neq 0$ $\chi(\omega,e)$ is generally asymmetric under the transformation $\omega \to -\omega$, as has been predicted for phase fluctuating superconductors \cite{Levin99}. Only in the zero charge limit is particle-hole symmetry restored (Fig.~\ref{asymmetry}).

\subsubsection{Case E: ``Small charge'' AdS$_2$ holographic superconductor}

The AdS-RN black hole at $T=0$ has a near-horizon $r \rightarrow r_+=(12)^{-1/4}\sqrt{\rho}$ limit that corresponds to the geometry of AdS${}_2 \times R^2$. This radial distance in AdS characterizes the energy-scale at which the CFT is probed, and one can show that fermionic spectral functions that have the same phenomenology as the strange metallic behavior observed in condensed matter systems arise from gravitational physics in this near horizon AdS${}_2$ region \cite{Faulkner09}. It is therefore of interest at which temperature the superconducting instability sets in.

\begin{figure}
\centering
\includegraphics[width=8cm]{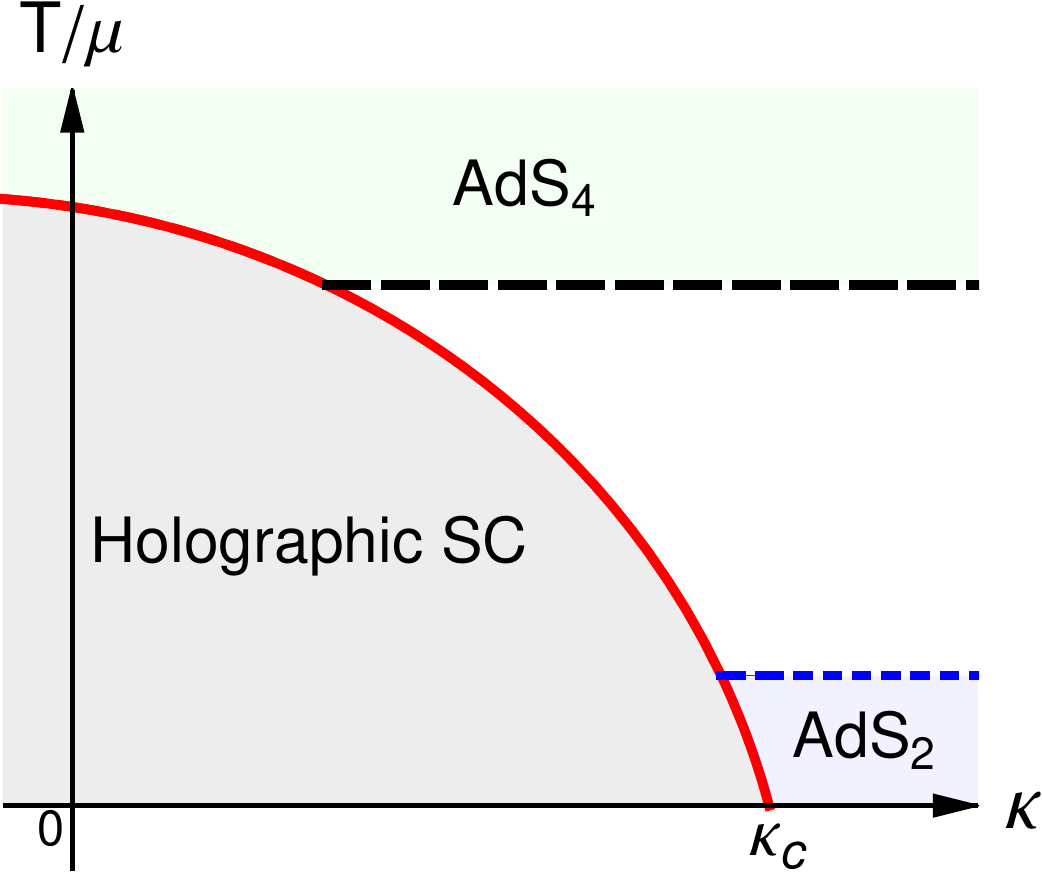}
\caption{
\label{cartoon}
 (Color online) {\bf A phase diagram of holographic superconductor including a double-trace deformation with strength $\boldsymbol{\kappa}$.} For $\kappa=0$ one has the minimal holographic superconductor, case D, where $T_c \sim \mu$. Increasing the value of $\kappa$ can decrease the critical temperature all the way to $T_c=0$ if one includes a non-minimal coupling to the AdS-gauge field (see text). The shaded regions indicate which region of the geometry primarily determines  the susceptibility. It shows that one must turn on a double-trace coupling to describe superconductors whose susceptibility is determined by AdS${}_{2}$-type physics. This is of interest as AdS${}_{2}$-type physics contains fermion spectral functions that are close to what is found experimentally.}
\end{figure}

In the case D simplest ``large-charge'' holographic superconductor all dimensionfull constants are of order one. Thus $T_c \sim \mu$ and the onset of superconductivity happens before one is essentially probing the near-horizon physics.\footnote{One can lower $T_c$ by lowering the charge of the order-parameter. Curiously the holographic theory even develops a Bose-condensate of a neutral order-parameter in the presence of a $U(1)$ chemical potential. Technically it remains difficult in this set-up to cleanly extract the AdS${}_2$ scaling. We have, however, kept the nomenclature of ``small charge'' superconductor indicating that $T_c \ll \mu$.  }
To access the AdS$_2$ near-horizon geometry we wish to tune $T_c$ as low as possible. 
This can be realized by  
combining a double trace deformation in the CFT with a non-minimal  ``dilaton-type'' coupling in the gravity theory\cite{Horowitz10}. When the order parameter $\mathcal{O}$ has scaling
dimension $\vartriangle_-<3/2$, $\mathcal{O}^\dagger\mathcal{O}$ is a relevant operator, and the IR of the field theory can be driven to a quantitatively different $T_c$/qualitatively different state by adding this relevant operator as a deformation 
\be\label{e:-looking2} S_{\mathrm{FT}}\to S_{\mathrm{FT}}-\int d^3x\tilde{\kappa}\mathcal{O}^\dagger\mathcal{O},\ee
where $\tilde{\kappa}=2(3-2\vartriangle_-)\kappa$. 
See Fig.~\ref{cartoon}.
This operation does not change the bulk action, but now we need to study the bulk gravitational theory using new boundary conditions for the scalar field. The retarded Green's function becomes\cite{Witten01} \be
\label{e:-small}
G_R \sim \frac{\psi_-}{\kappa\psi_--\psi_+}, 
\ee
and the susceptibility can be shown to take the Dyson-series RPA form:
\be 
\chi_{\kappa}= \frac{\chi_0}{1+\kappa \chi_0}.\ee

This already modifies $T_c$ but it can be further reduced by adding an extra ``dilaton-type'' coupling $|\Psi|^2F^2$ term to the minimal model action in equation~(\ref{action0}), \cite{Horowitz10}
\be
\label{e:-looking} 
\mathcal{S}_1=-\frac{\eta}{4}\int d^4x \sqrt{-g}|\Psi|^2F_{\mu\nu}F^{\mu\nu}.\ee 
In the normal phase, the AdS RN black hole, equation~(\ref{rnbh}), is still a solution to this action. The susceptibily again follows from the Green's function (\ref{e:-small}) in this background, which is built from solutions to the equation of motion for $\delta\Psi(r,x,y,t)|_{k=0}=\psi(r)e^{-i\omega t}$. With the two modifications (\ref{e:-looking2}) and
(\ref{e:-looking}) it equals
\be
\label{eompsid} \psi''+\bigg(\frac{f'}{f}+\frac{2}{r}\bigg)\psi'+\bigg(\frac{(\omega +eA_t)^2}{f^2}
+\frac{\eta\rho^2}{2r^4f}-\frac{m^2}{f}\bigg)\psi=0.
\ee

\section{Results and Discussions}

\begin{figure*}
\includegraphics[width=0.42\textwidth]{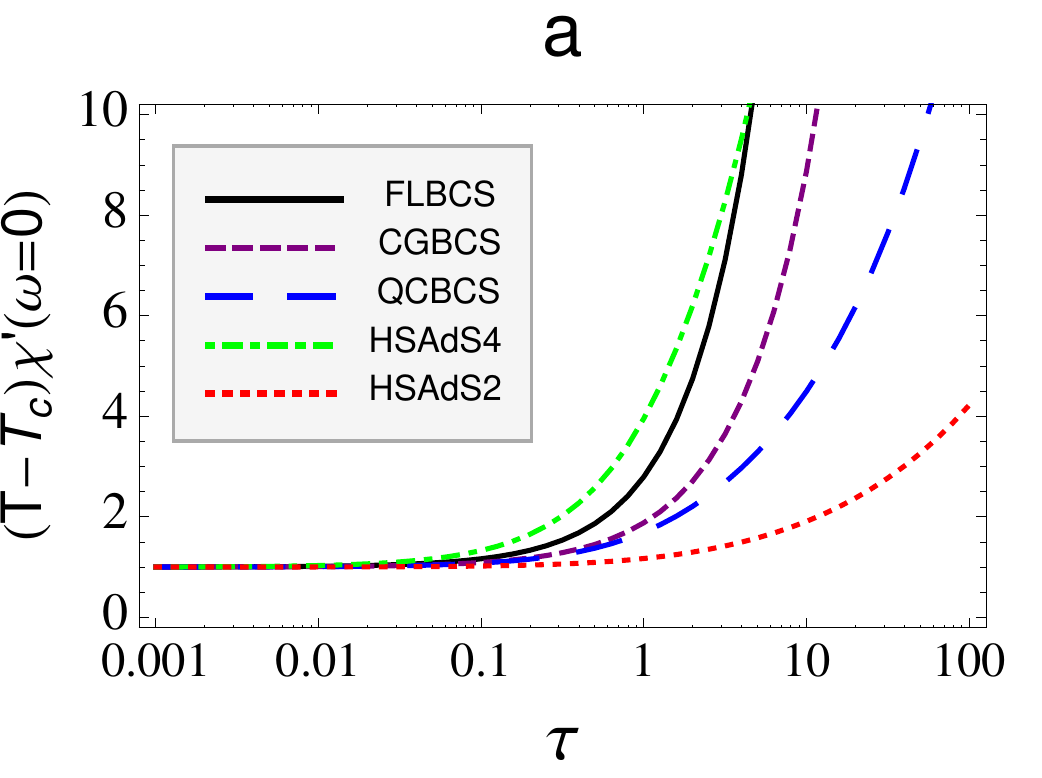} \quad
\includegraphics[width=0.42\textwidth]{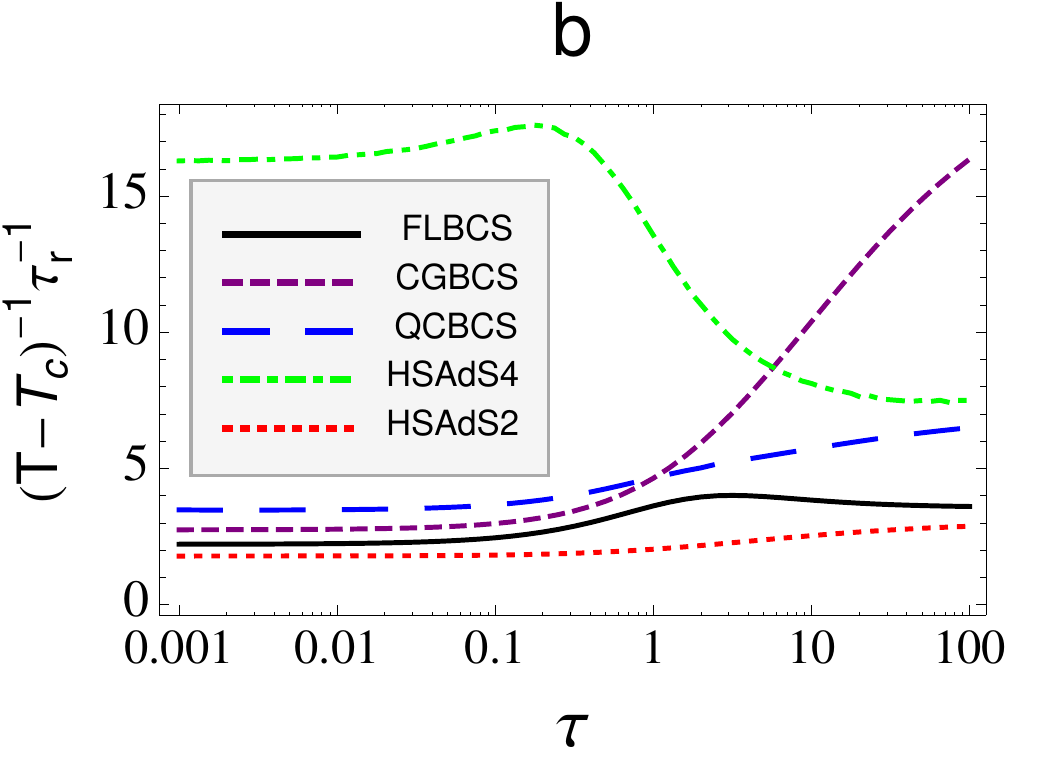}
 \caption{(Color online) \textbf{The universal mean field behavior of the pair susceptibility close to the superconducting phase
  transition.}  {\bf a}, Plot of the real part of the pair susceptibility at zero frequency rescaled by the distance to the 
  superconducting transition point,  i.e., $(T-T_c)\chi'(\omega=0,T)$, as function of reduced
  temperature $\tau=(T-T_c)/T_c$, for the five different models considered. 
  The horizontal axis is plotted on the logarithmic scale, and we use
  the normalization
  $(T-T_c)\chi'(\omega=0,T)\to1$ as $T\to T_c$. $\chi'(\omega=0)$ is a measure of
  the overall magnitude of the pair susceptibility in arbitrary units.
  $\chi'(\omega=0, \tau)$ can be determined from the experimentally measured imaginary part of the pair susceptibility by using the Kramers-Kronig 
  relation $\chi'(\omega=0,T)=\frac{1}{\pi}\int
  d\omega\chi''(\omega,T)/\omega$. {\bf b}, the inverse relaxation time $\tau_r$ rescaled by the distance to the 
  superconducting transition point, i.e., $(T-T_c)^{-1}\tau_r^{-1}$, as 
  function of reduced temperature $\tau$. The horizontal axis is also plotted on the logarithmic scale. The relaxation time is calculated from the relation 
  $\tau_{r}=[\partial \chi''/\partial\omega]_{\omega=0}/\chi'(\omega=0)$ (see text for equations). 
  In both plots, for all the five different models A-E, the curves become flat close to the transition temperature $T_c$ (here for $\tau\lesssim 0.1$), i.e., both
  $\chi'(\omega=0,T)$ and $\tau_r(T)$ behave as $1/(T-T_c)$,
  confirming the universal mean field behavior in this regime. We also see from {\bf
    b} that the ``large charge'' holographic superconductor (here with
  charge $e=5$) has a much shorter relaxation time than the ``small
  charge'' holographic superconductor (here with
  charge $e=0$).  
\label{fig:nearTc}}
\end{figure*}

\begin{figure*}
\includegraphics[width=0.32\textwidth]{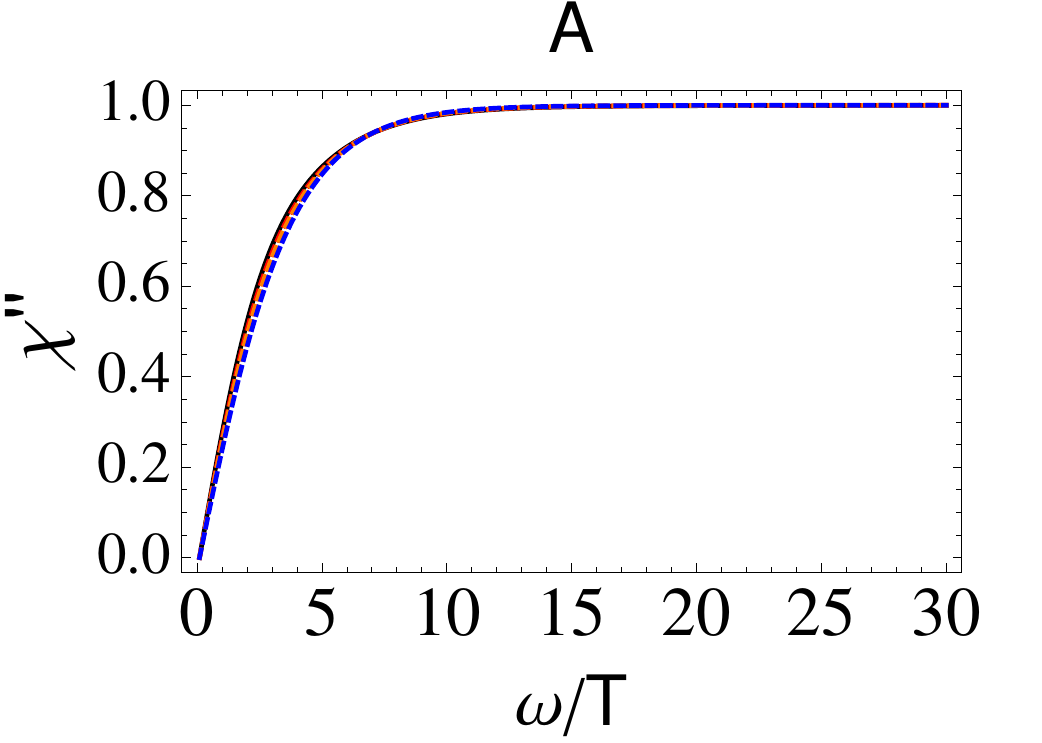}\quad
\includegraphics[width=0.32\textwidth]{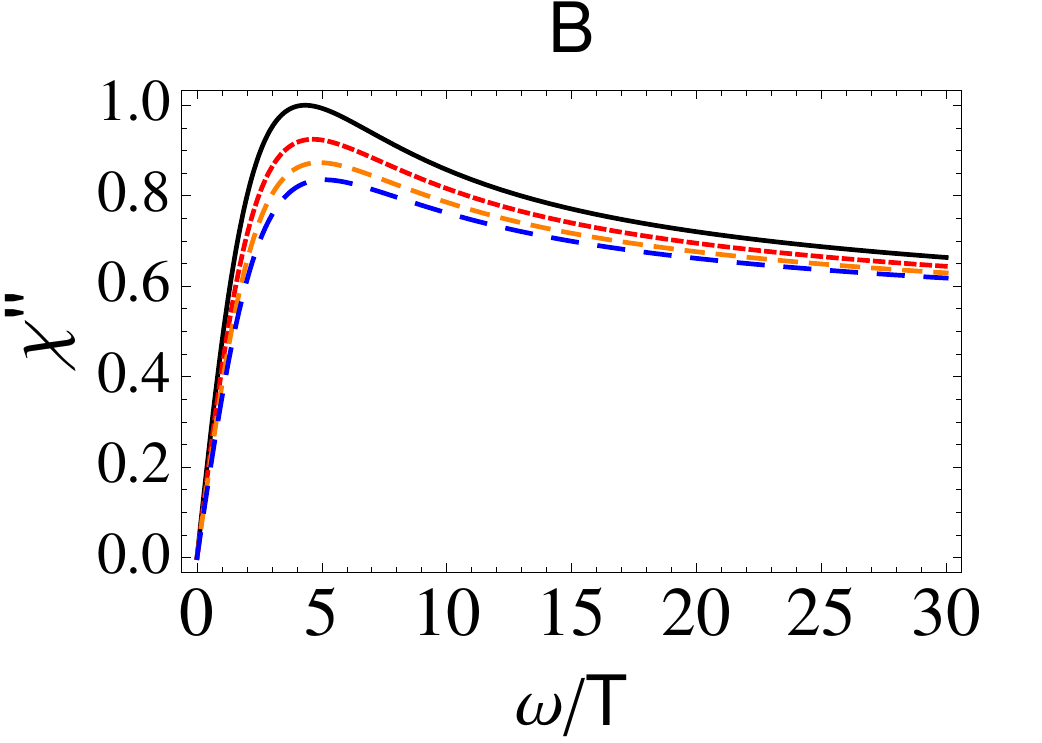}\\
\includegraphics[width=0.32\textwidth]{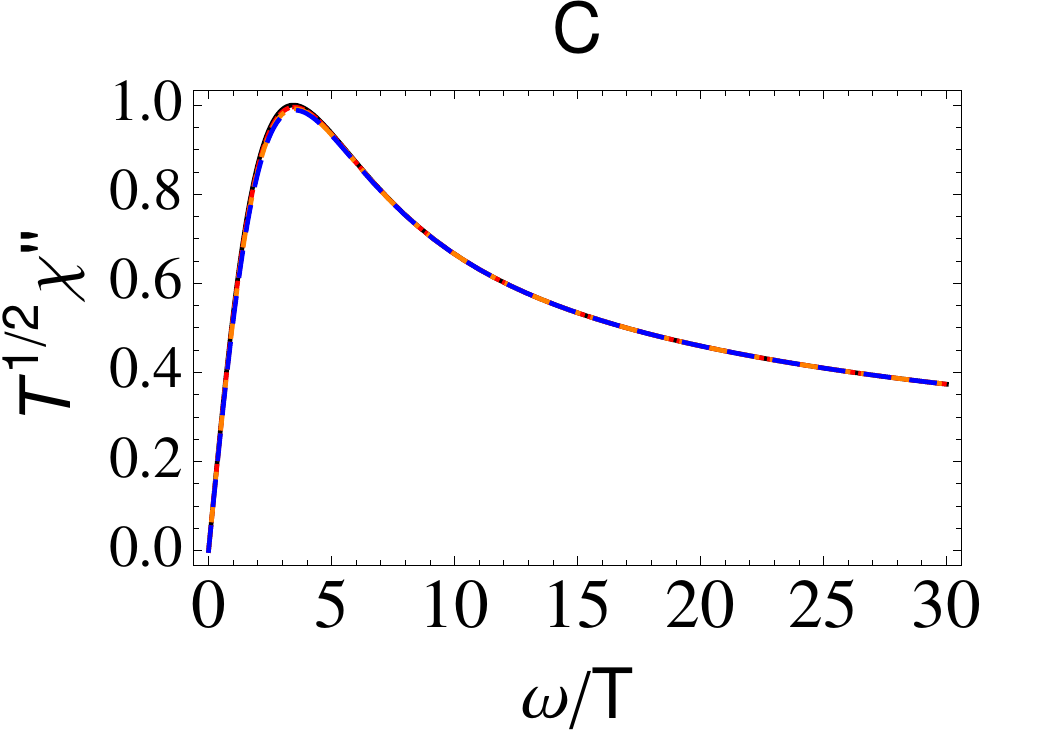}\quad
\includegraphics[width=0.32\textwidth]{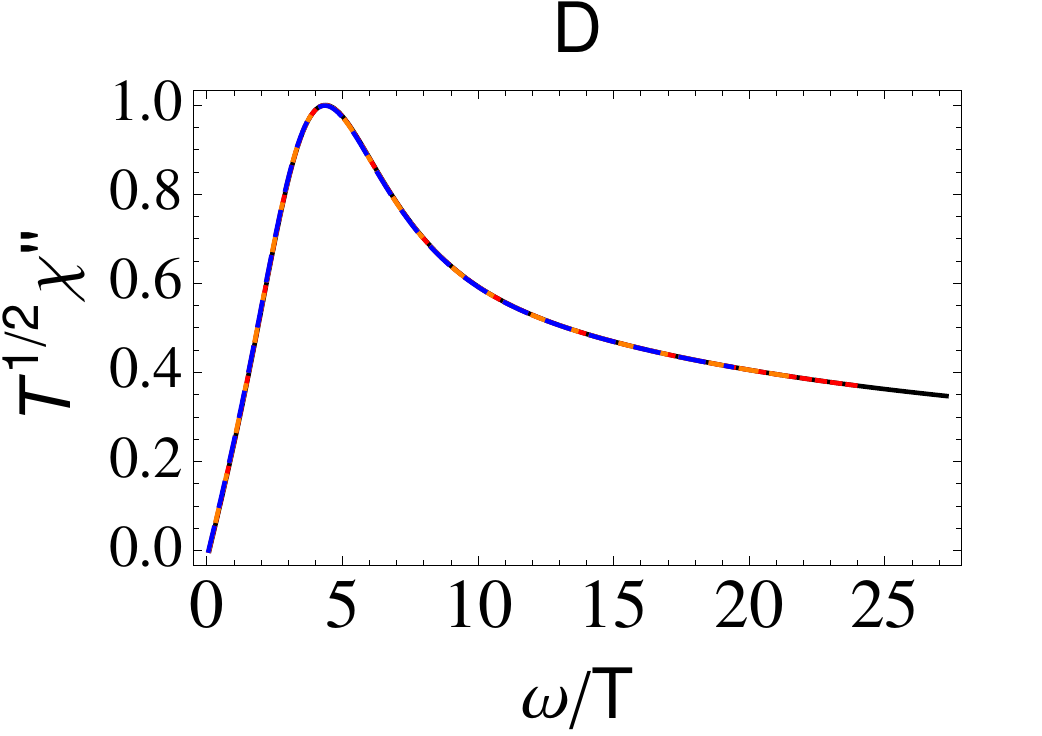}\quad
\includegraphics[width=0.32\textwidth]{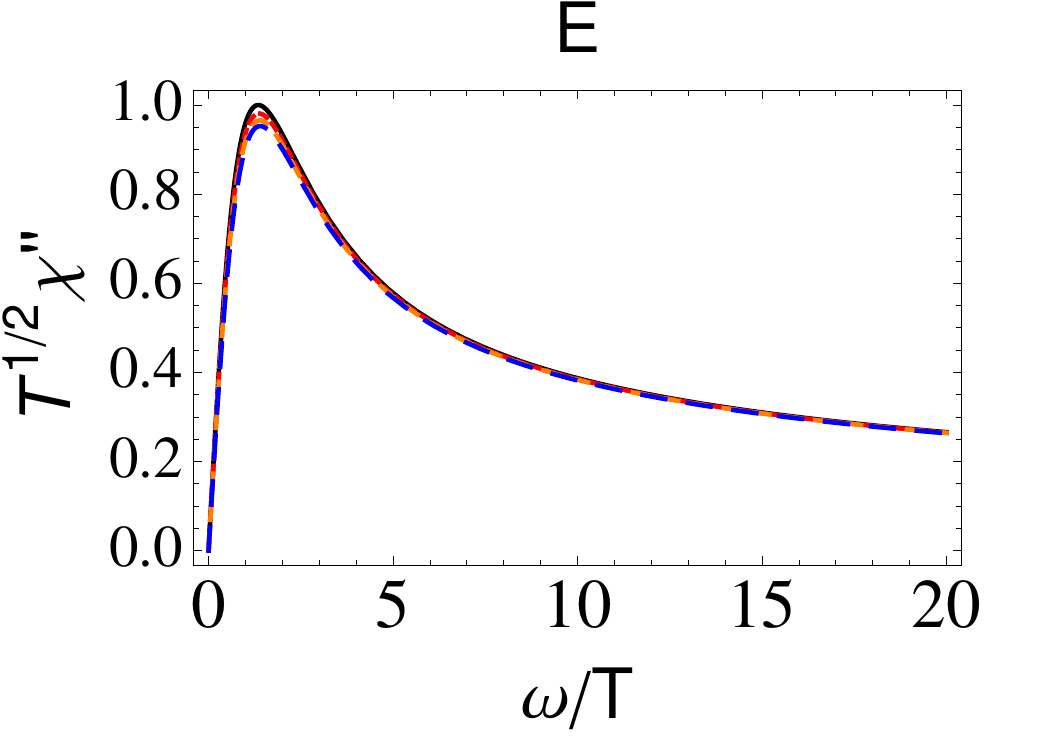}
\caption{(Color online) \textbf{Energy-temperature scaling line cuts. A-E,} High temperature line cuts of the imaginary part of the pair susceptibility rescaled by temperature to a certain power: $T^\delta \chi''$, as function of
  $\omega/T$ for the five different cases. Here in each figure we have plotted four different temperatures, with
   reduced temperatures $\tau=21,24,27,30$. As the vertical contour lines in Fig. 3 already revealed, the cases A, C, D and E exhibit a scaling collapse at
   high temperatures, whereas scaling collapse fails in
   the ``Hertz-Millis-Chubukov'' case B. Furthermore, 
   the line shape is quite different in cases C, D and E as compared to cases A and B.      
      For C, D and E, $\chi''(\omega, T)$ decays as power law at high 
     temperatures whereas for A and B $\chi''(\omega, T)$ approaches the Fermi liquid $\tanh$-form in the high temperature limit.
     The pronounced peak in cases C, D and E versus the flatness of A and B is signified by a plot of the full width at half maximum (see
     Fig. ~\ref{crossover}).} 
  
\label{fig:linecuts}
\end{figure*}

\begin{figure}[ht]
\begin{center}
\includegraphics[width=5cm, clip]{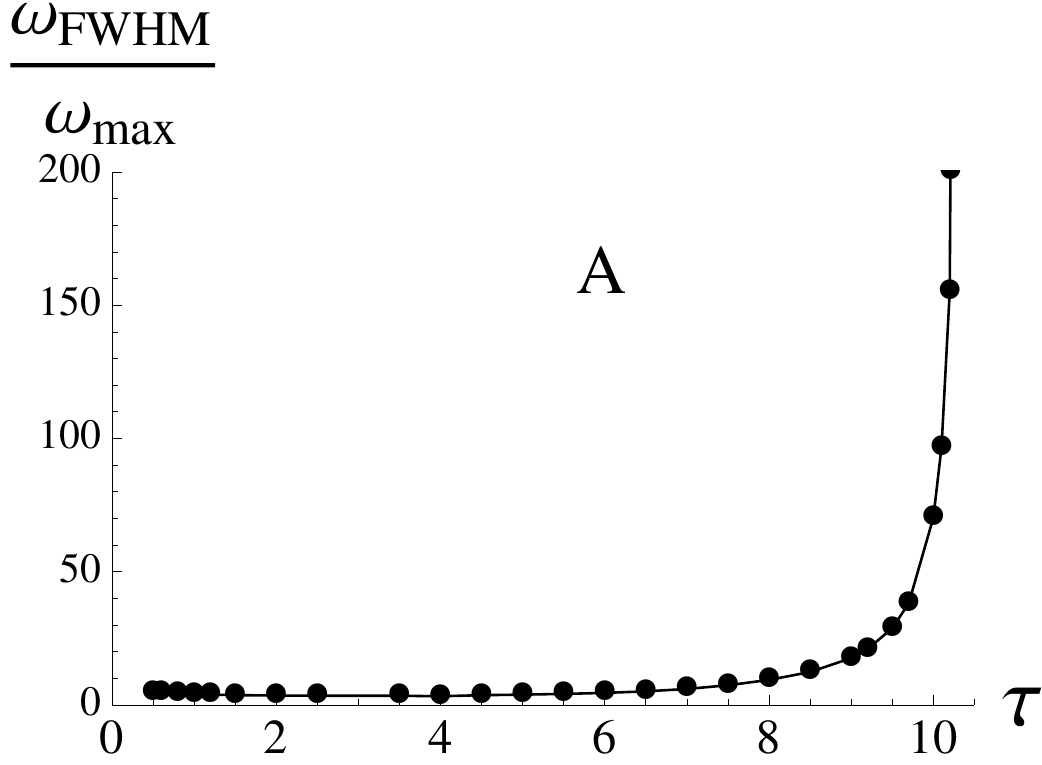}
\includegraphics[width=5cm, clip]{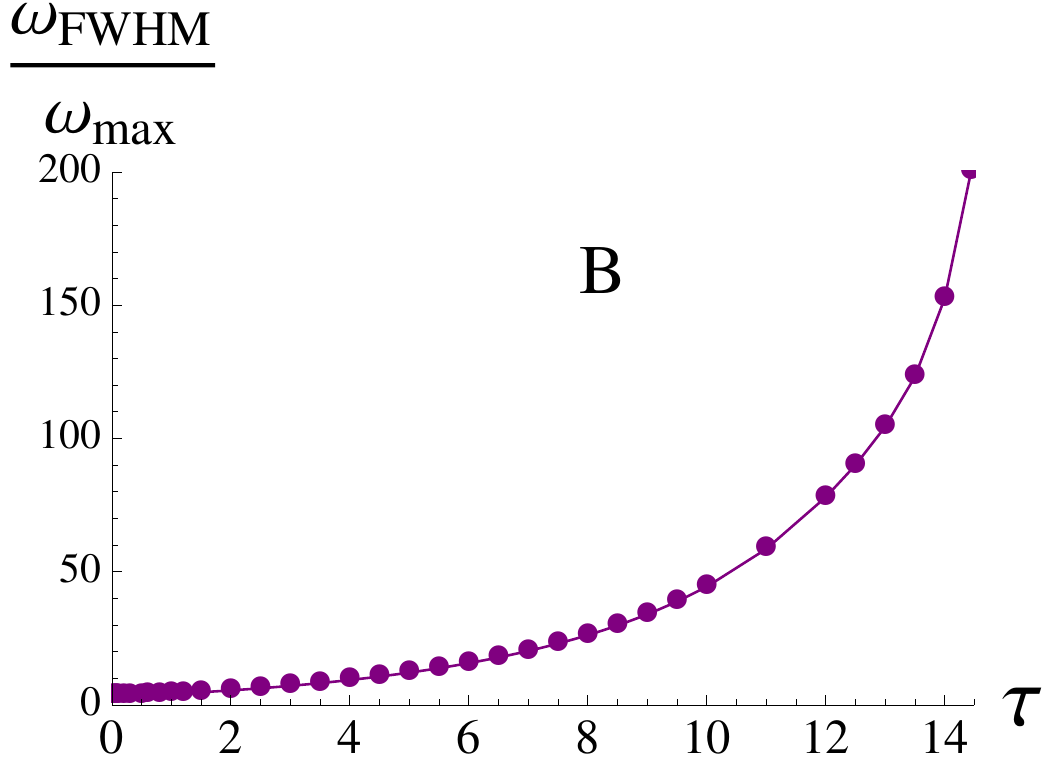}\\
\includegraphics[width=5cm, clip]{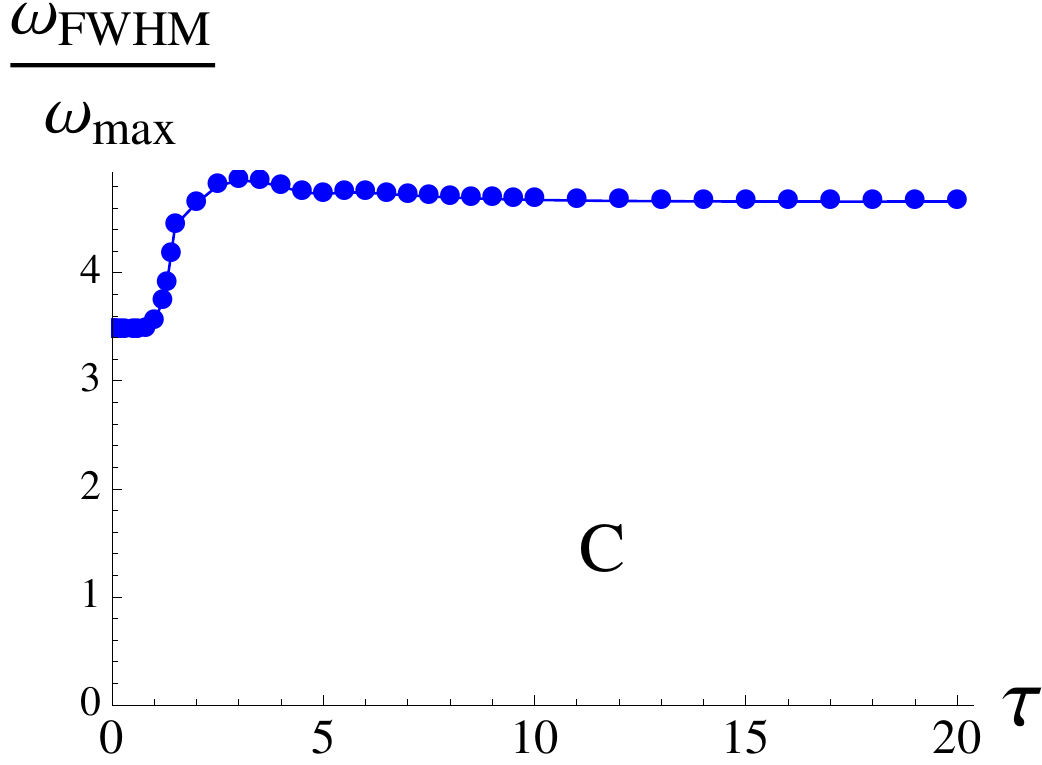}
\includegraphics[width=5cm, clip]{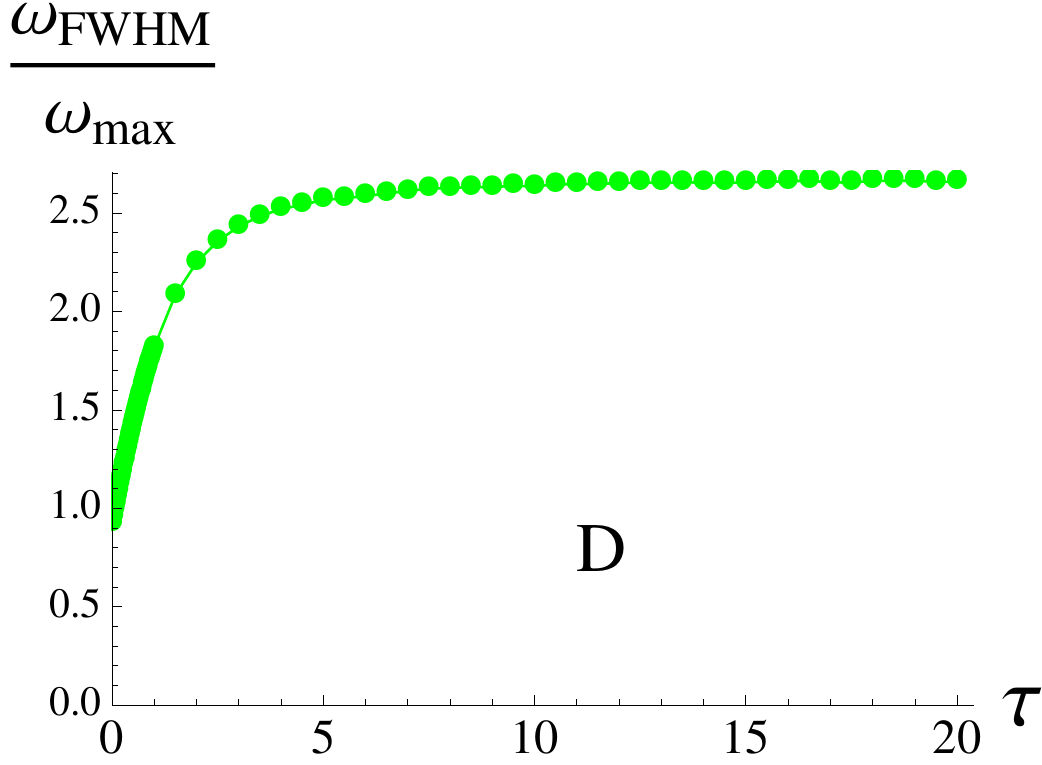}
\includegraphics[width=5cm, clip]{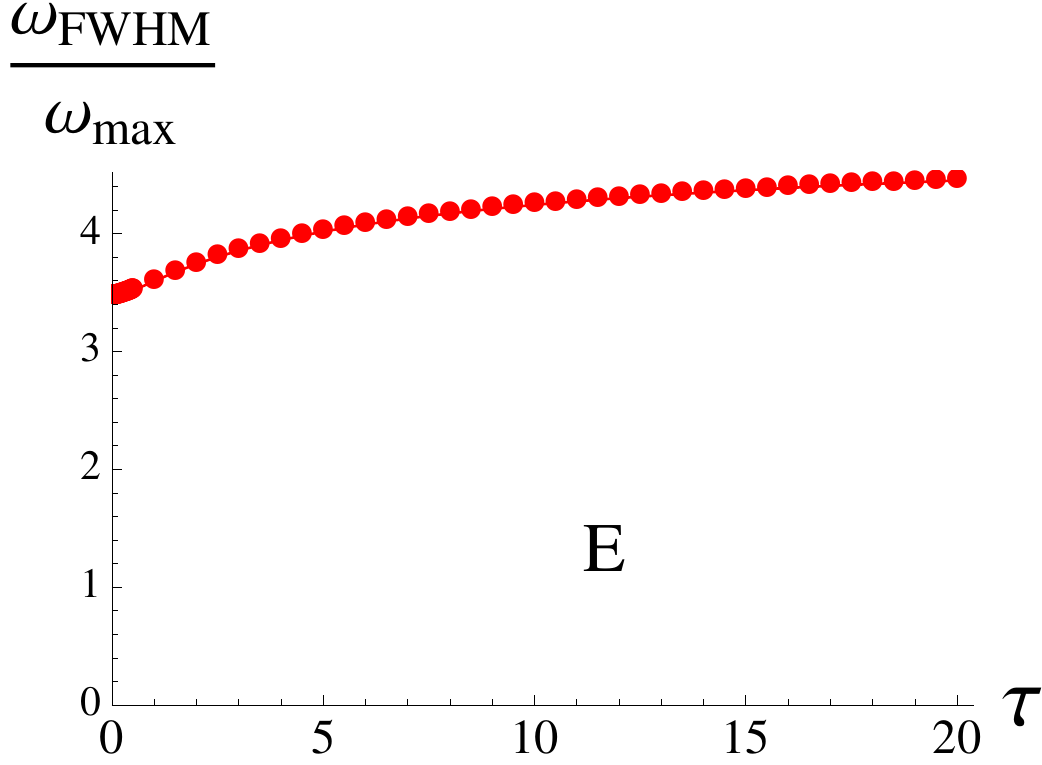}
\end{center}
\caption{(Color online) {\bf Peak width crossover.} Evolution of the relative peak width, i.e., the ratio of the full width at half maximum (FWHM) of the peak and peak location $\omega_\text{max}$, as a function of reduced temperature $\tau=(T-T_c)/T_c$ for the five different models. For FLBCS (A) and CGBCS (B), the ratio diverges at high temperature. For QCBCS (C) there is a sudden change from the low temperature relaxational behavior to the high temperature conformal field theory behavior. For the two holographic superconductors (D--E), the crossover from high temperature region to low temperature region is  more smooth.}
\label{crossover}
\end{figure}

Let us now explain why the experiment needs to cover a large range of temperatures and frequencies in order to extract the differences in physics. The thermal transition to the superconducting state
is in all cases a ``BCS-like'' mean field transition --- for A--C this is by construction, involving large coherence lengths, but for the holographic 
superconductors it is an outcome that is expected but not completely understood. 
As in all critical phenomena, 
the mean-field  universal behavior sufficiently close to the phase transition to the superconducting state is given by standard  
Ginzburg-Landau order parameter theory,
\begin{equation}
{\cal L} =  \frac{1}{\tau_r} \Psi \partial_t  \Psi + |\nabla \Psi|^2 +  i\frac{1} {\tau_{\mu}} \Psi \partial_t  \Psi  + \alpha_0(T-T_c) |\Psi|^2 + w |\Psi|^4 + \cdots  
\label{ginzlan}
\end{equation}
Evaluating the order parameter susceptibility in the normal state one finds,
\begin{equation}
\chi_\text{pair} (\omega, T) = \frac {\chi'_\text{pair}  (\omega=0,
  T)} { 1 - i \omega \tau_r - \omega \tau_{\mu}},
\label{eq:relaxpeak}
\end{equation}
Indeed in all cases Fig.~\ref{fig:nearTc}a shows the  familiar ``Curie-Weiss'' behavior $\chi'_\text{pair} (\omega =0,T) = 1 / [
\alpha_0 (T-T_c)]$, at temperatures $T_c\leq T \lesssim 3T_c$, with relaxation time $\tau_r\propto (T-T_c)^{-1}$. 
The time $\tau_{\mu}$ measures the breaking of the charge conjugation
symmetry at the transition. In the relaxational regime, the tunneling
current signal obtains the quasi-Lorentzian lineshape
$\chi''_\text{pair}(\omega)=\chi'(0)\tau_r \omega/[\tau_r
  ^2\omega^2+(1-\tau_\mu \omega)^2]$.  Since cases A--C are strongly retarded,
charge conjugation is effectively restored (i.e, $\tau_\mu=0$) for the usual reason that the density
of fermionic states is effectively constant (or symmetric, case C)  around $E_F$.  As for phase-fluctuating local pairs, the `strongly coupled' holographic 
superconductor D shows a quite charge-conjugation asymmetric result,
$\tau_\mu/\tau_r\approx 0.4$, while it is remarkable that the ``weakly coupled'' holographic case E 
is displays a near complete dynamical restoration of charge
conjugation ($\tau_\mu/\tau_r\approx 0$) (see Fig. ~\ref{asymmetry}).

In the Landau-Ginzburg regime the order parameter relaxation time $\tau_r$ does still give us a window on the underlying fundamental physics.
Strongly coupled quantum critical states are characterized by a fundamental ``Planckian'' relaxation time $\tau_{\hbar} = A \hbar / (k_B T)$ and the order parameter fluctuations in the normal state ought to submit to this universal relaxation. For rather elegant reasons 
this is the case in the
holographic superconductors (D,E) (see Appendix A). One finds that $\tau_r = A_{D/E} \hbar
 / [ k_B (T-T_c)]$, where 
$A_D\approx0.06$, $A_E\approx 1.1$ (
``zero temperature'' equals $T_c$ for the order parameter susceptibility). Not surprisingly this works in a very similar way for case C but viewed from 
this quantum critical angle the textbook BCS result that $\tau_r = (\pi/8) \hbar /[ k_B (T-T_c)]$ is rather astonishing. Although the underlying 
Fermi-liquid has a definite scale $E_F$ (e.g.,  its relaxation time is $\tau_\text{FL} = (E_F/k_BT) \tau_{\hbar}$) its pair channel is governed by effective 
conformal invariance, actually in tune with the quantum critical BCS moral.  

Given this ``quasi-universality'' near the phase transition, 
one has to look elsewhere to discern the pairing mechanism from the 
information in the pair susceptibility. It is obvious where to look: Fig.~\ref{fig:expsignal} shows that the differences appear at temperatures large compared to $T_c$ 
involving a large dynamical range in frequency. This is the challenge for the experimental realization. In this 
large dynamical range one distinguishes directly all quantum critical cases (B--E) for which the
contour lines in Fig.~\ref{fig:expsignal} acquire a convex shape, from simple BCS with 
fanning-out
contours. One sees the reasons for this more clearly in figures~\ref{fig:tempscaling}
and~\ref{fig:linecuts}, which plot $T^\delta \chi''_p(\omega/T,\tau)$, i.e., a
rescaling by temperature. Figure~\ref{fig:tempscaling}
displays the same temperature range as in figure~\ref{fig:expsignal}, figure~\ref{fig:linecuts} shows
several line-cuts at high temperatures. In the simple BCS case A the high temperature pair susceptibility 
is just the free Fermi gas result $\chi'' (\omega,T) = (1/E_F) \tanh ( \omega / 4T)$, linearly increasing with frequency initially and becoming constant for 
$\omega > 8 T$.  In cases B--E the pair susceptibility deep in the normal state increases with decreasing frequency down to a scale set by 
temperature to eventually go to zero linearly at small frequency as required by hydrodynamics. The observation of such a behavior would reveal 
a significant clue regarding a non conventional origin of the superconductivity. The frequency independence of $\chi''_\text{BCS}(\omega)$ reveals the 
``marginal'' scaling that is equivalent to  the logarithmic singularity in $\chi' (\omega =0)$ that governs the BCS instability. In constrast, the critical temperature peak  in $\chi'' (\omega)$ in cases B--E reveals a ``relevant'' scaling behavior in the pair channel: a 
stronger, algebraic singularity is at work  giving  away that the quantum critical electron system is intrinsically supporting a more robust superconductivity
than the Fermi gas.

The observation of such a peak implies that one can abandon the search for some ``superglue'' that enforces pairing in the
Fermi gas at a ``high'' temperature. Instead the central question becomes: what is the origin of the relevant scaling flow in the pair channel 
in the normal state, and is the normal state truly quantum critical in the sense of being controlled by conformal invariance? Fig.~4 shows that, if it is, the pair susceptibility must display energy-temperature 
scaling in this high temperature regime. 
Both the quantum critical BCS (C)  and the two holographic cases (D,E) embark from the 
assumption that the high temperature metal is governed by a strongly interacting quantum critical state that is subjected to the hyper-scaling underlying 
the energy-temperature scaling collapse. Specifically the pair operator itself is asserted to have well-defined scaling properties, as
in 1+1-dimensional Luttinger liquid. Such ``truly'' quantum
critical metals have no relation whatever with the Fermi liquid, but this is not quite the 
case for the Hertz style critical glue case (B). Although the fully re-summed Eliashberg  treatment of  the strongly coupled ``singular glue'' 
$\lambda (\omega) \sim 1/\omega^{\gamma}$ completely changes the pair susceptibility  relative to the simple BCS case it is still a perturbative theory 
around the Fermi-liquid. It remembers that it is based on a  Fermi liquid with a characteristic scale $E_F$ and this prohibits the
energy temperature scaling, as illustrated in Figs~\ref{fig:tempscaling}B,~\ref{fig:linecuts}B.   
   
The observation of energy-temperature scaling in the high temperature  pair susceptibility would therefore reveal the existence of a true non-Fermi
liquid quantum critical state formed from fermions. Although it remains to be seen whether it has any bearing on the condensed matter systems, the
only controlled mathematical theory that is available right now to deal with such states of matter is the AdS/CFT correspondence of string theory. 
It has its limitations: the ``bottom up'' or ``phenomenological'' approach of relevance in the condensed matter context  should be regarded as
a generalized scaling theory, which reveals generic renormalization flows associated  with strongly interacting quantum critical states encountered in the presence
of fermions at a finite density. However, the scaling dimensions, rates of relevant flows and so forth associated with a particular theory/universality class
are undetermined in this bottom up approach. Cases D and E are two limiting cases of such generic RG flows.  The ``large charge'' case D departs from a ``primordial''  Lorentz invariant critical state at zero density (encoded in an
AdS$_4$ geometry in the gravitational dual)  that is natural 
in supersymmetric quantum field theory while it is far fetched as a UV theory for condensed matter systems. 
A better
holographic contender for condensed matter physics is case E. Here
the holographic superconductivity 
is
governed by the emergent  quantum criticality 
associated with the near horizon AdS$_2$ geometry of the extremal Reissner-Nordstrom black hole. 
This is dual
to 
an (unstable) infrared fixed
point 
where
the normal state shows
the traits of the marginal Fermi liquid \cite{Faulkner10}. 

For the pair susceptibilities the distinction between case D and E is only quantitative at zero momentum, associated with a choice of different scaling dimensions. 
The crucial difference is with the other scenarios. In the holographic cases no ``external glue''
 is at work.
The superconducting instability is an intrinsic property associated
with the strongly coupled fermionic critical matter.
As can be seen directly from the pair susceptibilities in Figs. \ref{fig:tempscaling}D,
 \ref{fig:tempscaling}E, the superconducting 
 correlations builds up through a very smooth but 
 rapid flow from the conformal high temperature regime to the
 relaxational regime associated with the thermal transition.
 The smoothness of the flow towards the instability is also emphasized when one considers more closely
 the way that the relaxational peak morphs into the conformal peak as
 function of temperature: 
in the QCBCS case this can be relatively sudden given 
 that scale is introduced through the characteristic glue energy, while
in both the AdS$_4$ and AdS$_2$ cases this is just a very smooth cross
over flow (see Fig. \ref{crossover}). 

\section{Conclusions}

In this paper we demonstrated, through explicit calculations of the
pairing susceptibility in five different models, the existence of sharp qualitative
difference between the truly quantum critical models --- phenomenological QCBCS or the holographic models --- and the Hertz-Millis type
models with respect to energy-temperature scaling. In the Hertz-Millis
type models for pairing, the pairing channel is assumed to be secondary. The single
particle Green's functions, and/or certain bosonic 
quantities in the particle-hole
channel, e.g. magnetic susceptibilities, are considered to be primary and carry the criticality, enjoying energy-temperature
scaling. The pairing susceptibility is
assumed to be a derived quantity, and it remains sensitive to the underlying Fermi-energy. Thus generally one does not expect to have
energy-temperature scaling in the pairing channel,
or at the best scaling can only occur with extreme fine tuning.
 The essence of QCBCS and the holographic approach is to take the superconducting order parameter itself to be a conformal field in the quantum critical region. 
This is the underlying reason for energy-temperature scaling in these
models. The observation of energy-temperature scaling with an obviously nonzero scaling exponent in the pairing susceptibility would unambiguously reveal the non-BCS nature of the pairing mechanism and the non-Hertz-Millis nature of the quantum critical state.
The contrast between superconductivity emerging from a strongly interacting 
fermionic quantum critical state and any mechanism that sets out from a Fermi-liquid is qualitatively so different that the proposed
experiment might finally settle the basic rules associated with superconductivity in quantum critical systems.

\section*{Acknowledgements}

 We would like to acknowledge Jan Aarts, Mihailo $\rm \check{C}$ubrovi$\rm \acute{c}$, J. C. Seamus Davis, Dimitrios Galanakis, Sean A. Hartnoll, Hans
Hilgenkamp, Mark Jarrell, Sergei I. Mukhin, Andrei Parnachev, Catherine Pepin, Jan van Ruitenbeek and Jian-Xin Zhu for
stimulating discussions. This research was supported in part by a VIDI Innovative Research
Incentive Grant (K. Schalm) from the Netherlands Organisation for
Scientific Research (NWO), a Spinoza Award (J. Zaanen) from the
Netherlands Organisation for Scientific Research (NWO) and the Dutch
Foundation for Fundamental Research on Matter (FOM).      

\section*{Appendix A: Relaxational behavior in holographic
  superconductors: near-far matching}

\setcounter{equation}{0}
\renewcommand{\theequation}{A\arabic{equation}}

\def\cG{{\cal G}}
A remarkable aspect of the AdS/CFT computation is that the relaxational behavior is directly encoded in the geometry. Near $\omega \to 0$ an analytic expression for the Green's function follows from a near-horizon/AdS-boundary matching method first used in ref. ~\onlinecite{Faulkner09}.
The result is that for $\omega \to 0$, the Green's function is of the form
\be\label{gr2} {G}_R(\omega ,T, e)\sim\frac{b_+^{(0)}+b_+^{(1)}\omega  +\mathcal{O}(\omega ^2)+\cG(\omega ,T)\big(b_-^{(0)}
+b_-^{(1)}\omega +\mathcal{O}(\omega ^2)\big)}{a_+^{(0)}+a_+^{(1)}\omega +\mathcal{O}(\omega ^2)+\cG(\omega ,T)\big(a_-^{(0)}+a_-^{(1)}\omega 
+\mathcal{O}(\omega ^2)\big)},
\ee
where $\cG(\omega ,T)$ is the near-horizon ``IR-CFT'' Green's function defined in a similar way as the full ``AdS-CFT'' Green's function, i.e., it is the ratio of leading and subleading coefficients at the boundary of the near-horizon region of a solution to the equation of motion. The coefficients $a_\pm^{(n)}(e, T), b_\pm^{(n)}(e, T)$ are determined by matching this IR-solution to the  ``UV''-solution near the AdS-boundary at spatial infinity. They can only be obtained numerically. Note that when $e=0$ particle-hole symmetry dictates that in that case $a_\pm^{(1)}=b_\pm^{(1)}=0$.

As $G_R$ is the Green's function for the order parameter it must develop a pole at $\omega=0$ for $T=T_c$. Thus when $T \to T_c, \omega \to 0$, the Green's function, equation~(\ref{gr2}), takes the  form 
\be\label{gr2tc} 
G_R(\omega ,T,e)\sim \frac{\gamma_0}{\beta_0(T-T_c)+ i\omega \beta_1 +\omega \beta_2},\ee
where $\gamma_0=b_+^{(0)}(e, T_c), ~\beta_0=\partial_Ta_+^{(0)}(e, T_c),~\beta_1=\lim_{\omega \rightarrow 0} \frac{1}{i\omega} \cG(\omega, T_c) a_-^{(0)}(e, T_c)$ and $\beta_2=a_+^{(1)}(e, T_c)$ 
Comparing to the universal relaxational behavior of the susceptibility (equation (6) in the main text)
\be
\chi(\omega, T) = \frac{\chi'(\omega=0,T)}{1-i\omega \tau_r -\omega \tau_\mu}
\ee
we recognize the Curie-Weiss susceptibility $\chi'(\omega=0,T)=\frac{\gamma_0}{\beta_0(T-T_c)}$, and the particle-hole asymmetry parameter $\tau_\mu  =- \frac{\beta_2}{\beta_0(T-T_c)}$, which indeed vanishes when $e=0$. But most interestingly the relaxation time
\be
\tau_r =\lim_{\omega \rightarrow 0} \frac{i}{\omega\beta_0(T-T_c)} \cG(\omega, T_c) a_-^{(0)}(e, T_c)
\ee
is directly given in terms of the IR Green's function $\cG(\omega, T)$. The AdS gravity response function therefore directly knows about the relaxational dynamics in the dual conformal field theory.

There are essentially two different regimes of interest
\begin{enumerate}
\item For $\omega \ll T$,  the near-horizon IR Green's function
 takes the universal form $\cG(\omega, T) = -i\omega/4\pi T$. Thus $\tau_r = \alpha_0/(T-T_c)$. This is gravity version of the universal relaxation that for $\omega \ll T$,
  $\chi'' ={\rm Im} G_R $
should always be linear in $\omega$. (This frequency regime applies to both case D and E.)
\item 
For  $T\ll\omega\ll\sqrt{\rho}$, the IR Green's function is completely determined by the $SO(1,2)$ conformal symmetry of the near-horizon $T \simeq 0$ AdS$_2$ region. As a consequence the IR Green's function must be a power law in frequency:
\be
\cG\sim\omega^{\delta_+-\delta_-},
\ee
where $\delta_{\pm}$ are the two possible {IR}-conformal dimensions of the scalar field controlled by its dynamics the AdS$_2$ geometry, and we focus on real $\delta_{\pm}$.\cite{Faulkner09} In terms of the parameters explained in the Case E subsection (see equation (\ref{eompsid})) these conformal dimensions are
\be \delta_\pm=\frac{1}{2}\pm \sqrt{\frac{1}{4}+2r_+^4m^2-4r_+^4e^2-\eta},
\ee
For this range of frequencies the susceptibility will therefore also exhibit scaling but with non-``Curie-Weiss'' exponents: $\chi'' \sim \omega^{-\delta_++\delta_-}$. (This frequency regime only applies to case E. For case D the temperature $T_c \simeq \mu \sim \sqrt{\rho}$ and $T$ cannot be much smaller than $\sqrt{\rho}$ in the normal state.)
\end{enumerate}
For $\omega\gg\sqrt{\rho}$, one is outside of the regime of validity of ~(\ref{gr2tc}).  Indeed there is no ``relaxation'' for such high frequencies. Instead the Green's function is now determined by the UV-theory and  all temperature/chemical potential effects are subleading. In this case the UV-theory is a 2+1 dimensional CFT dual to AdS${}_4$ and the two-point correlation function is completely fixed by the $SO(3,2)$ symmetry  $\chi''\sim 1/\omega^{2 \nu}$ where $\nu=\frac{1}{2}\sqrt{{9+4m^2}}$. (Recall from equation (\ref{eompsi}) that we are using ``alternate quantization''. For ``standard quantization'' one would have $\chi''\sim \omega^{2 \nu}$.)






\section*{Appendix B: Pairing with Marginal Fermi liquid}

\setcounter{equation}{0}
\renewcommand{\theequation}{B\arabic{equation}}

To illustrate how powerful and universally distinctive the qualitative differences in energy-temperature scaling are, we study
here the pair susceptibility of the Marginal Fermi liquid (MFL) which
has been a prime candidate for some strange metallic states. A further motivation is that recent numerical calculations
\cite{Jarrell10, Jarrell11} found quantum critical scaling for a MFL when combined with a van Hove singularity (vHS). 

We first calculate 
the pairing susceptibility built out of a fermion bubble with MFL self-energy and a smooth density of states (DOS). Both a smooth BCS type pairing 
glue and a quantum critical glue are considered. We find that for both types of pairing interactions, energy-temperature scaling is severely broken, exhibiting clear 
distinction with QCBCS and the holographic approach. When now combined with an extended van Hove singularity, MFL can produce
 a ``quasi-conformal'' pair susceptibility \cite{Jarrell10, Jarrell11}, but extreme fine tuning is required. The vHS has to be precisely at the 
 Fermi-energy and in the whole frequency range the density of states has to exactly have a power law dependence 
 on frequency in order to give rise to a pair susceptibility that is subject to a perfect energy-temperature scaling. 
 By detuning the vHS away from the Fermi energy, or incorporating another scale even at the boundary of the measured frequency range, this scaling is lost.

\subsection*{MFL and MFL+vHS pair susceptibilities}

 In real frequency, the imaginary part of the MFL self energy is of the form
\begin{equation}
 \Sigma(\omega)=-a\pi \left\{
\begin{array}{ll}  {\rm max}\left\{|\omega|, T\right\}, & \mbox{for $\omega<\omega_E
$} \\  \omega_E, & \mbox{for $\omega>\omega_E$}. \end{array} \right.
 \end{equation} 
From the spectral representation, 
\begin{equation}
 \Sigma(i\omega_n)=\frac{1}{\pi}\int_{-\infty}^{\infty}d\nu\frac{\Sigma''(\nu)}{\nu-i\omega_n},
 \end{equation}
we obtain the self energy in imaginary frequency
\begin{equation}
 \Sigma(i\omega_n)=-a\omega_n\left( \frac{2T}{\omega_n}\arctan \frac{T}{\omega_n}- \frac{2\omega_E}{\omega_n}\arctan \frac{T}{\omega_n}+\pi\frac{\omega_E}{|\omega_n|}+\log\frac{\omega_n^2+\omega_E^2}{\omega_n^2+T^2}\right).
 \end{equation}

\begin{figure*}
\includegraphics[width=0.35\textwidth]{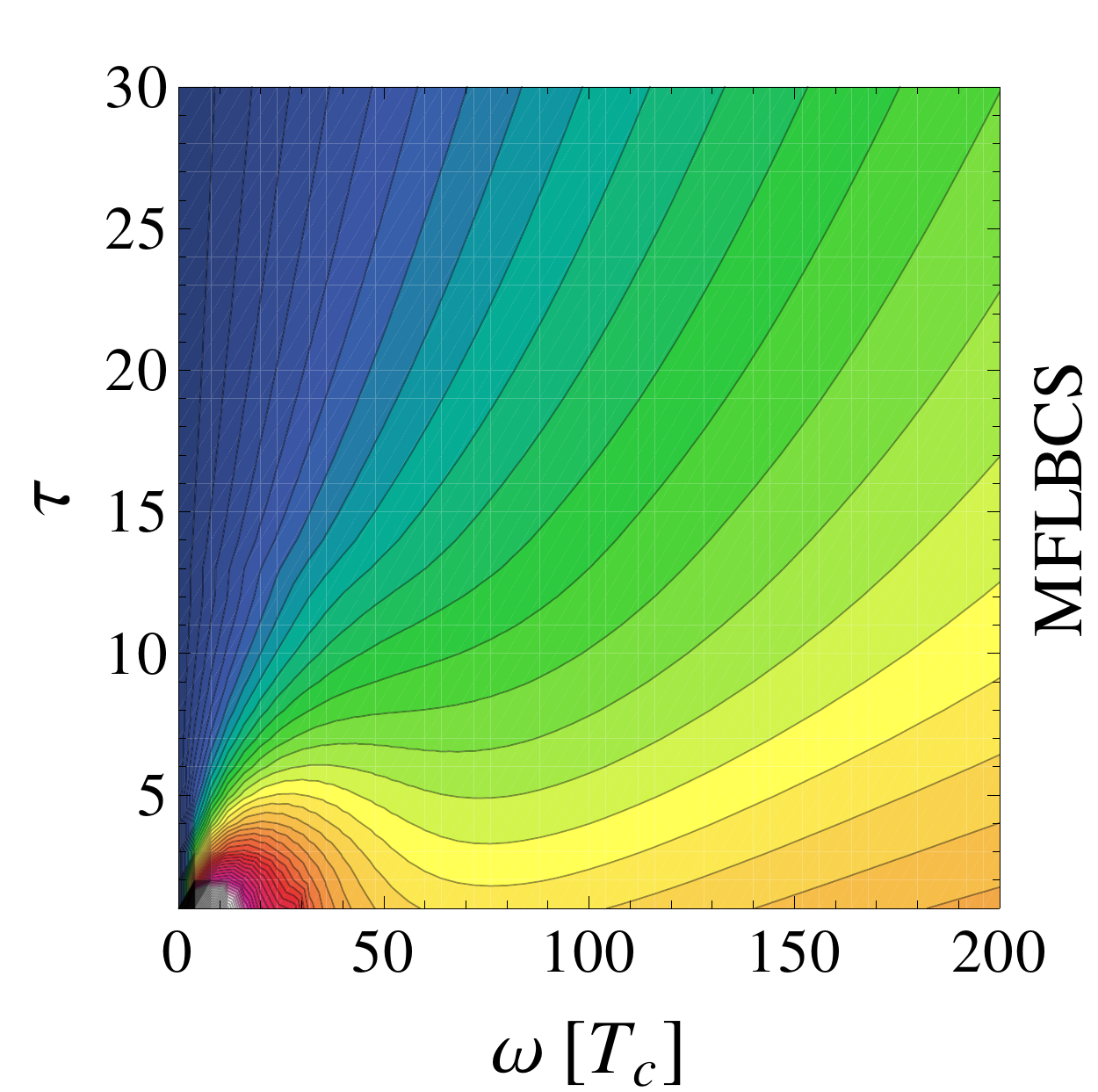}
\includegraphics[width=0.35\textwidth]{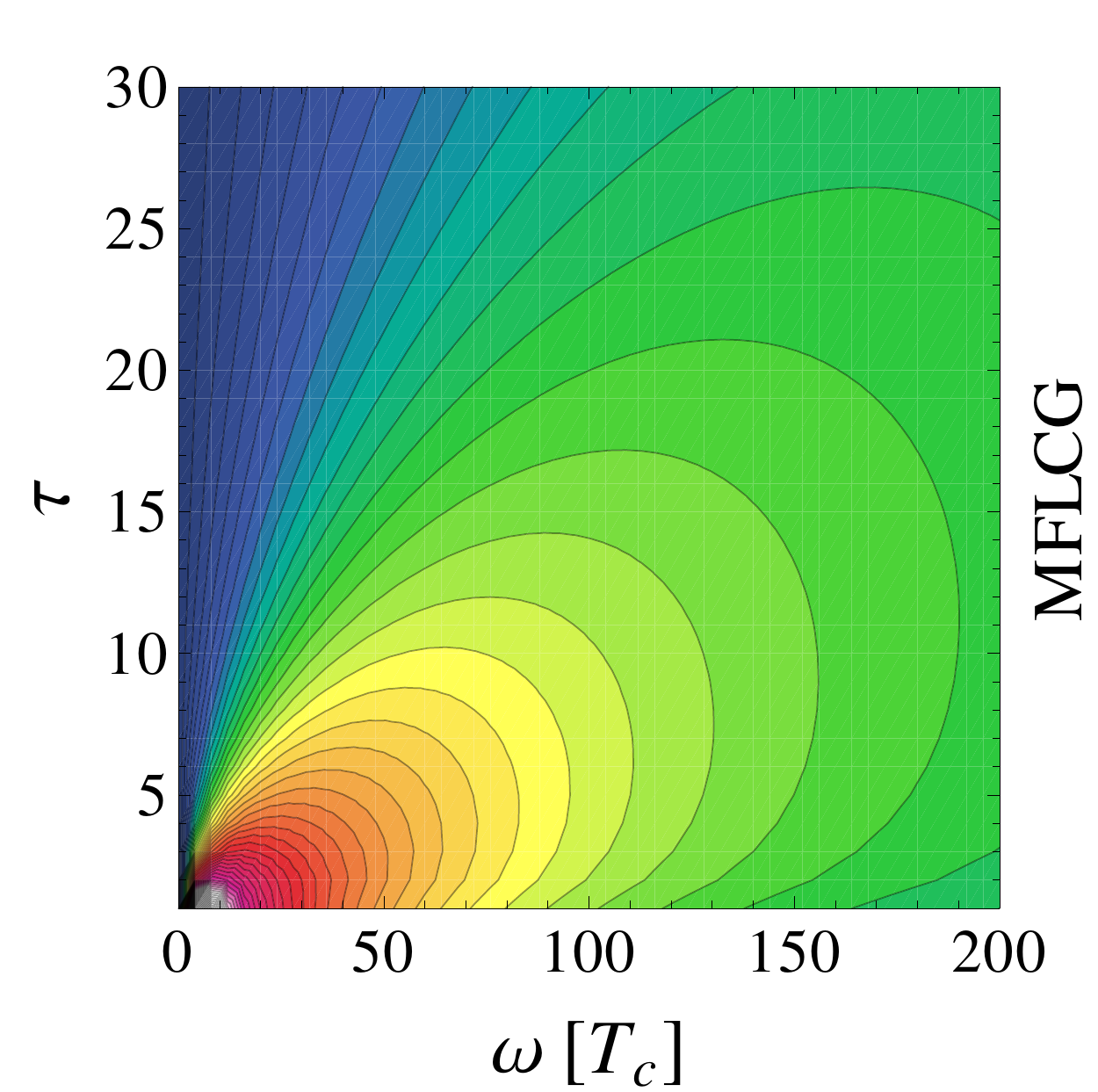}
\includegraphics[width=0.35\textwidth]{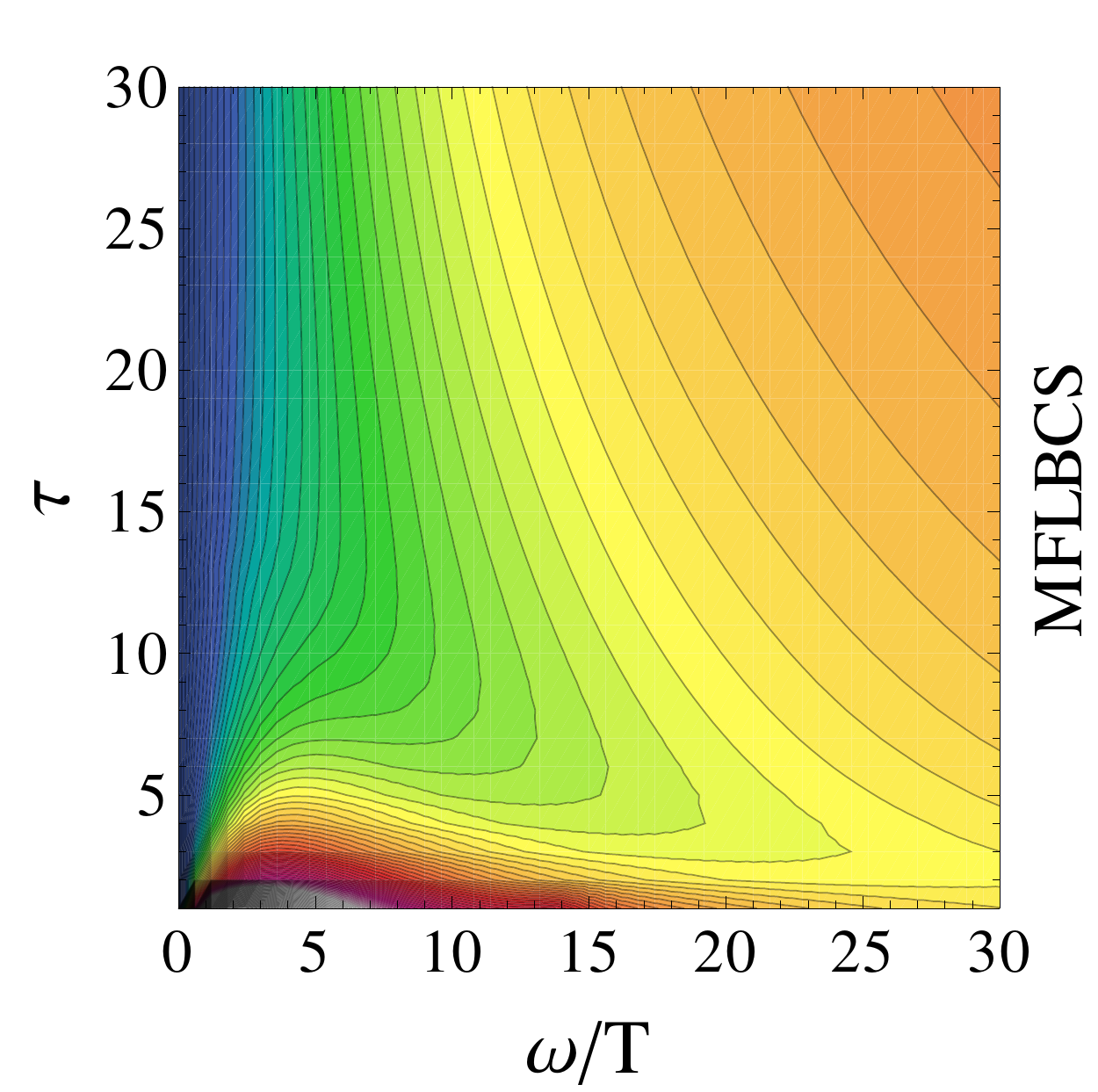}
\includegraphics[width=0.35\textwidth]{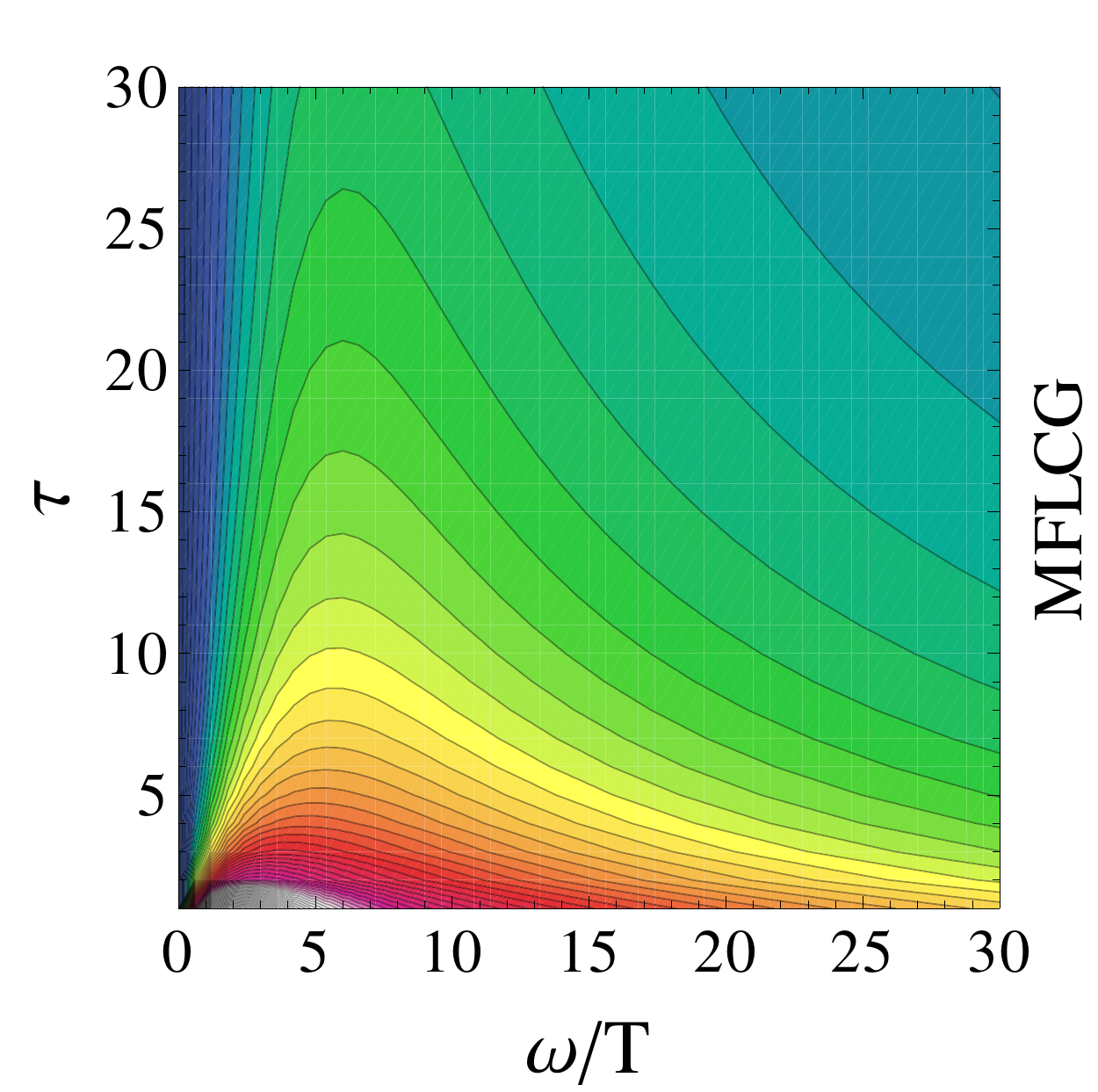}
\caption{(Color online) {\bf Marginal Fermi liquid pair susceptibility with smooth
    density of states.} {\it Top}: False-color plot of the imaginary part of the pair susceptibility $\chi''$ as function of frequency
  $\omega$ (in units of $T_c$) and reduced temperature $\tau=(T-T_c)/T_c$, for two different models:  
  marginal Fermi-liquid with BCS pairing and marginal Fermi-liquid with critical glue. In both cases, the density of states is taken to be constant.
  {\it Bottom}: the same plot, but now the horizontal axis is rescaled by temperature while the magnitude is rescaled by
temperature to a certain power: we are plotting $T^{\delta}\chi''(\omega/T,\tau)$, in order to show energy-temperature scaling at high temperatures. 
Here for both models $T_c=0.01$ and $\delta=0$. The color scheme is the same as used in the main text. For MFLBCS, the parameters are $a=0.3, \omega_E=1, g=0.9627, \omega_b=0.5$. 
For MFLCG, we take $a=0.4, \omega_E=0.2, \gamma=1/3, \Omega_0=0.0134$.
\label{fig:MFL1}}
\end{figure*}

Consider first the case where the DOS is a constant of
energy. We calculate $\chi''(\omega, T)$ with both a smooth BCS type
pairing glue (MFLBCS) and a quantum critical glue (MFLCG). 
The results are plotted in Fig.~\ref{fig:MFL1}, from which we see no
energy-temperature scaling for either of the two models. In addition, one can check that at large
temperatures for MFLBCS, $\chi''$ goes over to the BCS $\tanh$
form. The pair susceptibility is thus still marginal in this
sense. The inclusion of a nontrivial self energy destroys the ``marginal'' scaling behavior
of FLBCS. For MFLCG,
 the effects of the glue interactions are so strong that one ends up with a
 result that is barely distinguishable from CGBCS.

To our knowledge, the only way that the pair susceptibility of a MFL
can resemble that of QCBCS/HSAdS$_2$ in some sense is to invoke a van
Hove singularity in the spectrum.
The idea that the presence of vHS in the DOS is responsible for high temperature superconductivity has been around
for some time (see \cite{Friedel89, Markiewicz91, Newns92, Abrikosov93} and references therein).
 An extended van Hove singularity right at the Fermi level can
 produce a relevant pair susceptibility, i.e. the real part of the
 pair susceptibility $\chi'(\omega=0,T)$ has
 an algebraic temperature dependence \cite{Jarrell11}. But as will be shown below, extreme
 fine-tuning is needed to get energy-temperature scaling for the
 imaginary part of the pair susceptibility. Moreover, although there
 are indications of the presence of extended vHS in
 cuprates \cite{Shen93, Campuzano94}, to the best of our knowledge
 they have not been found in
 typical heavy fermion materials.

With the inclusion of a nontrivial DOS $N(\epsilon)$, the electronic vertex operator becomes
\begin{equation}
 \Gamma_0(i\nu_n, i\Omega)=\frac{T}{N_0} \int_{-\infty}^{\infty}d\epsilon N(\epsilon)\frac{1}{-i\nu_n-\epsilon- \Sigma(-i\nu_n)}\frac{1}{i\nu_n+i\Omega-\epsilon- \Sigma(i\nu_n+i\Omega)}.
\end{equation}
For MFL, $\chi''(\omega,T)$ can only be calculated numerically. But
the basic picture can be illustrated by considering the
non-interacting limit, where one has simply
$\chi_0''(\omega)=N(\omega/2)\tanh(\omega/4T)$. In this case, one can
easily see that, to get energy-temperature scaling for $\chi''$,
i.e. $\chi''(\omega,T)\to T^{\delta}{\cal F}(\omega/T)$, the DOS has to be a power of energy in the whole frequency range that is
experimentally relevant, $N(\epsilon)=
|\epsilon|^{-\alpha}$. Any deviation from this special form will break
energy-temperature scaling. This can be illustrated by considering
several explicitly deformations from the strict power form of the DOS.

\begin{figure*}
\includegraphics[width=0.3\textwidth]{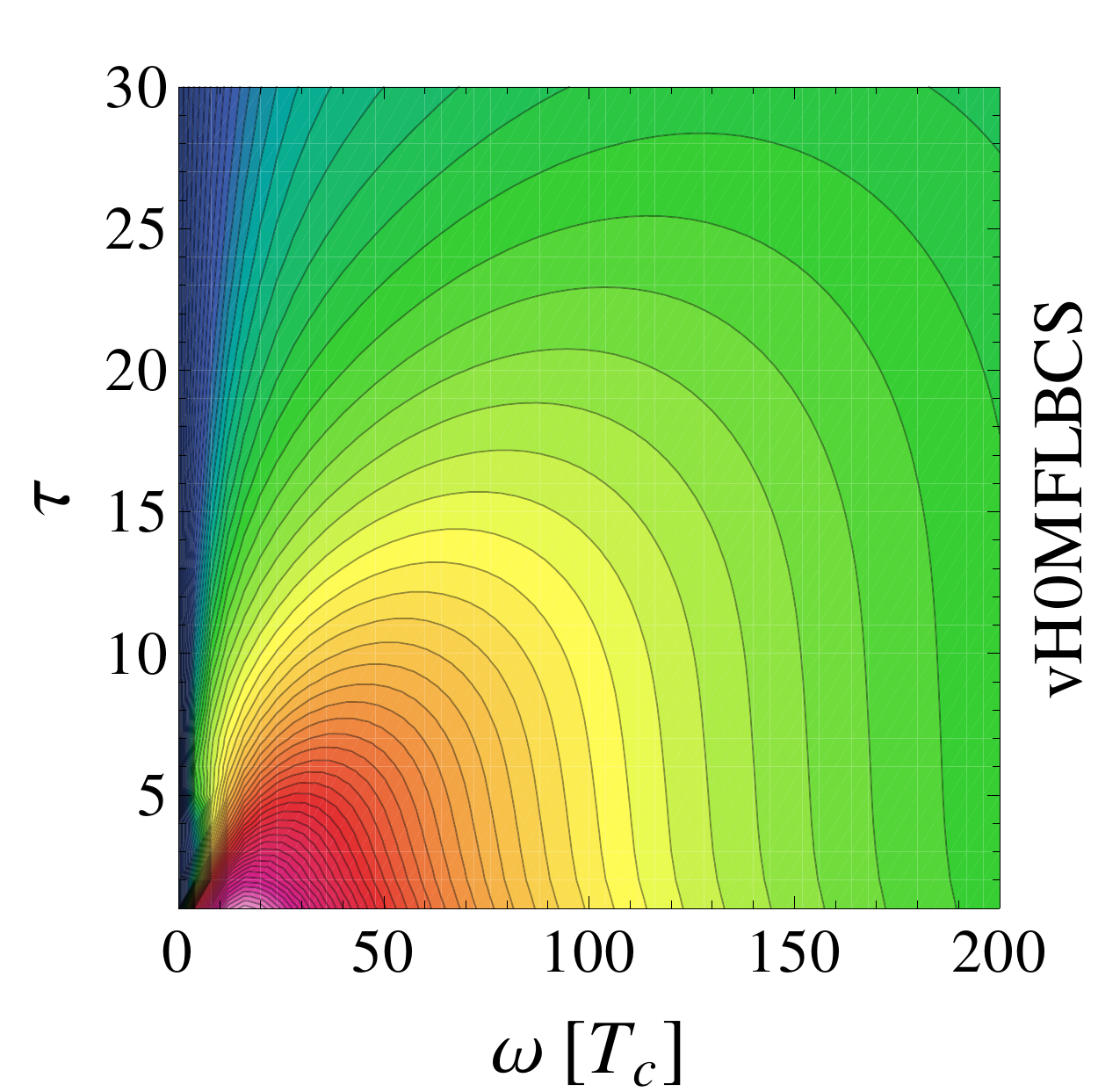}
\includegraphics[width=0.3\textwidth]{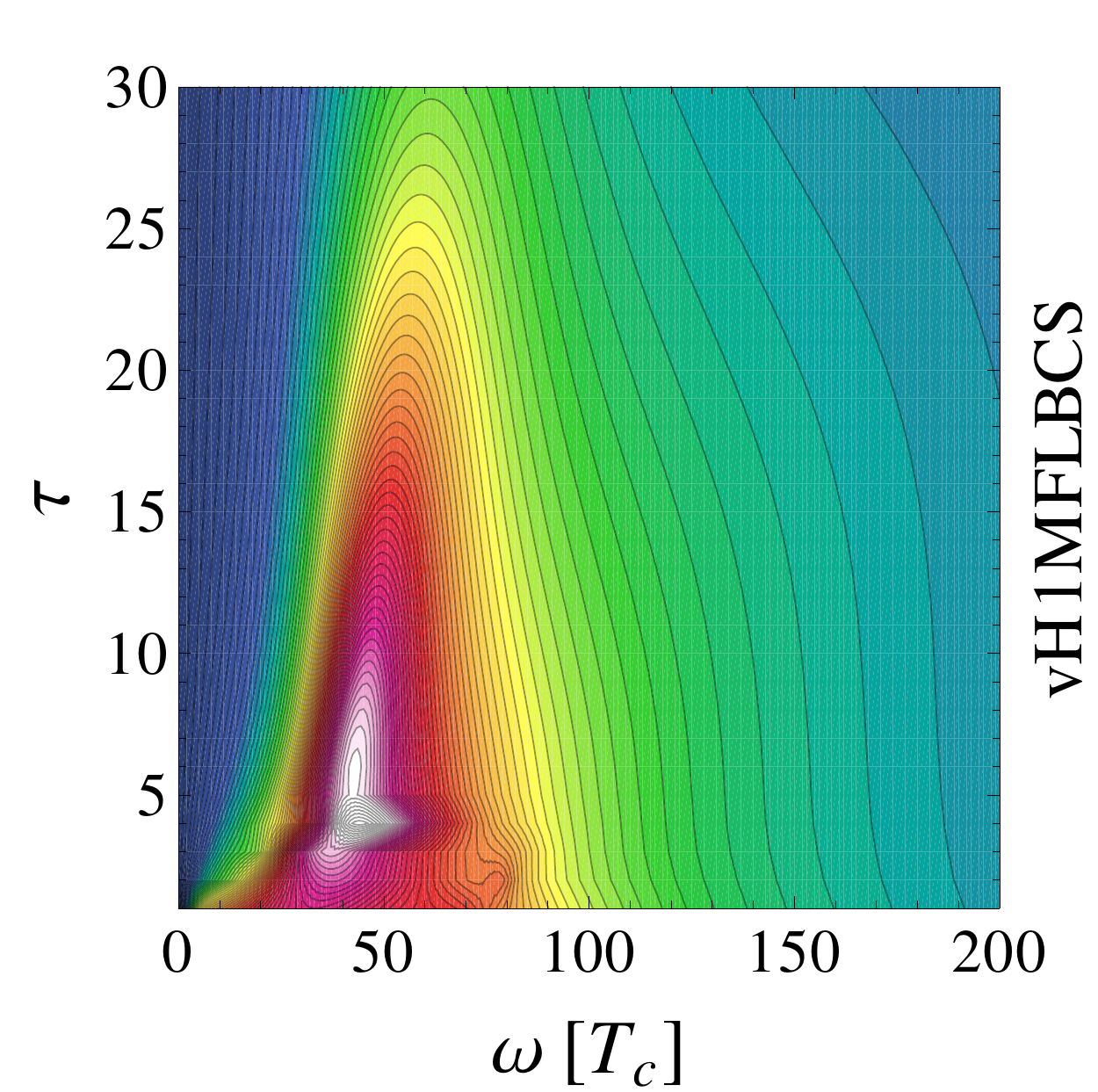}
\includegraphics[width=0.3\textwidth]{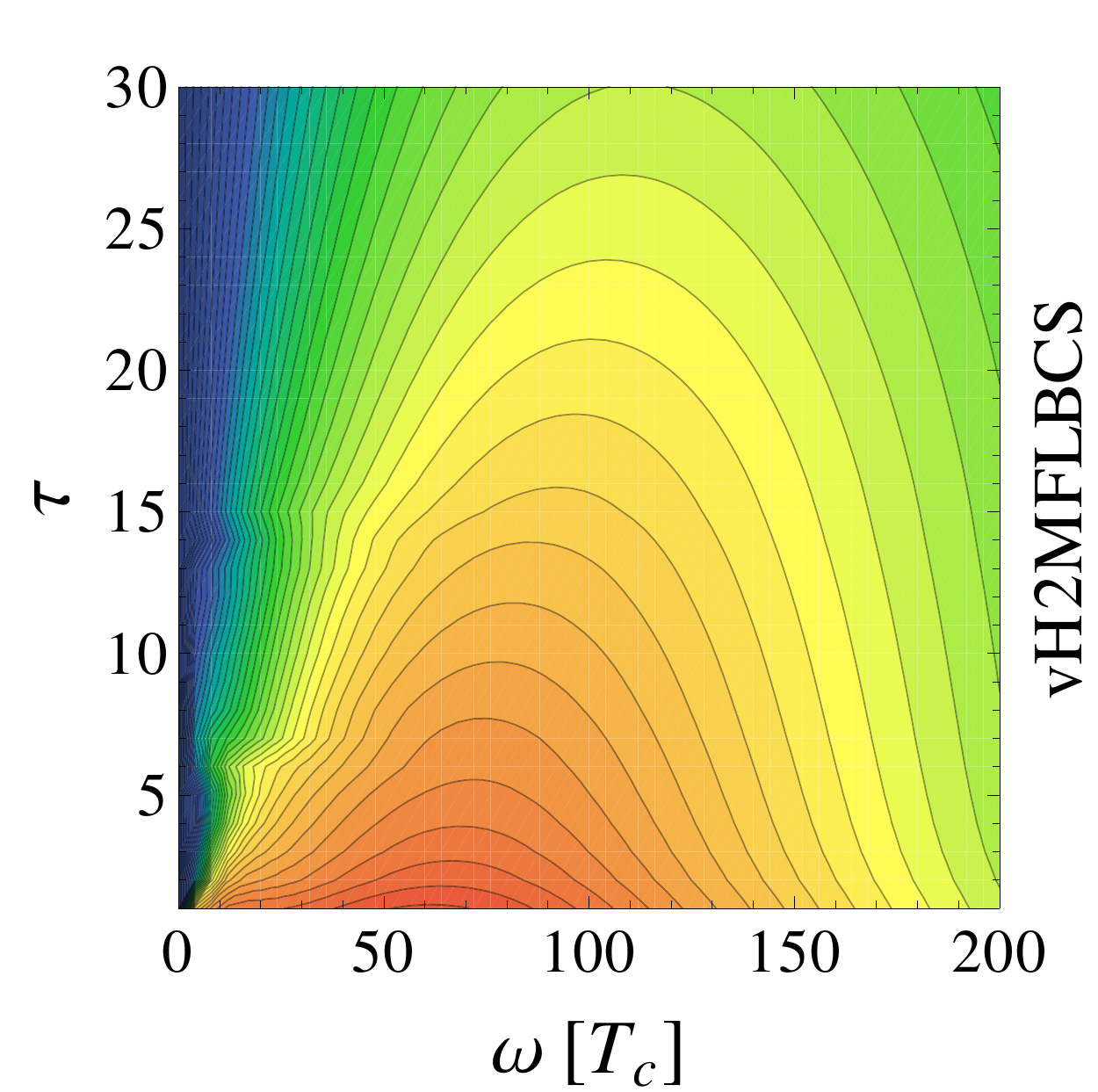}
\includegraphics[width=0.3\textwidth]{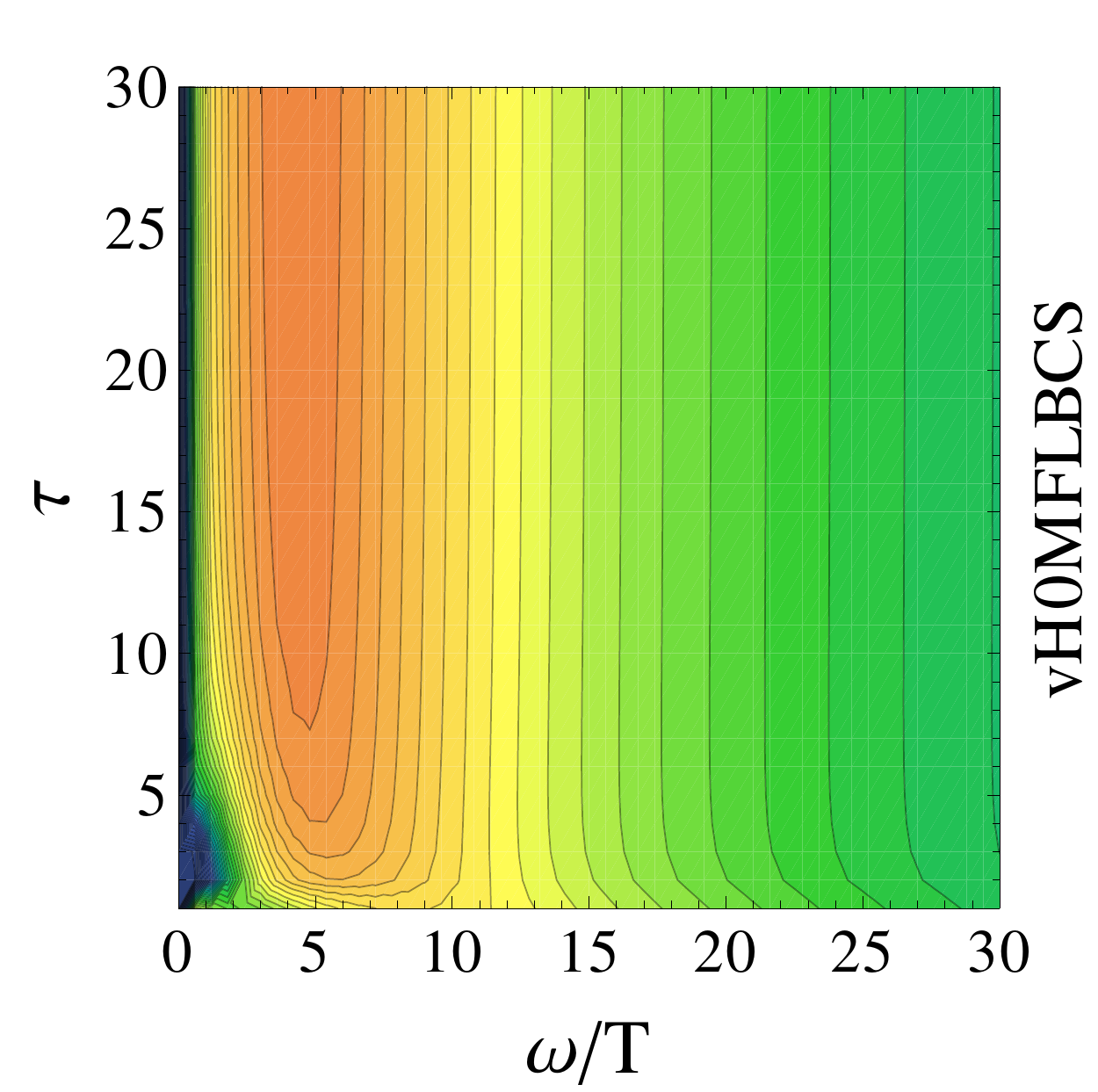}
\includegraphics[width=0.3\textwidth]{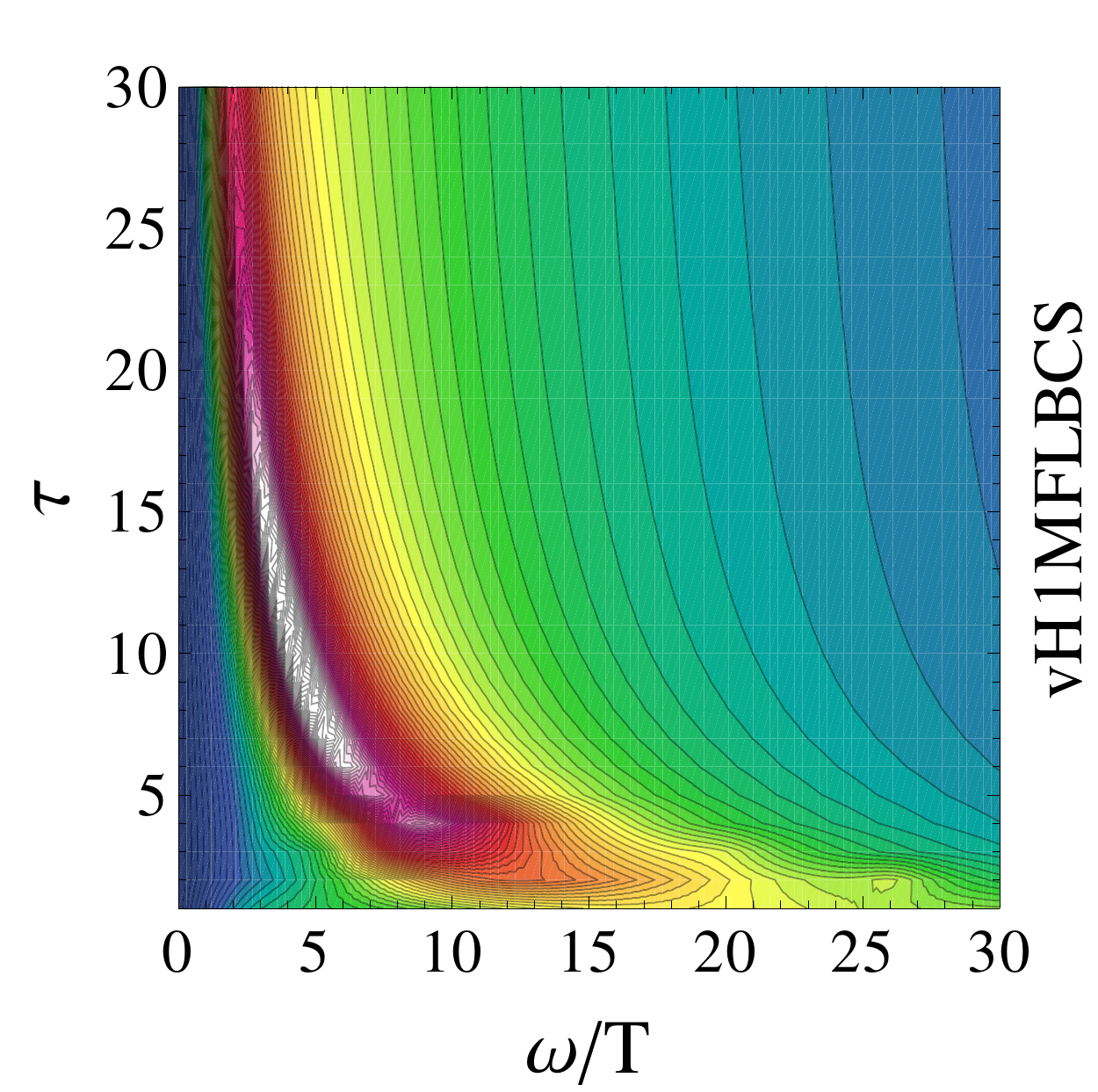}
\includegraphics[width=0.3\textwidth]{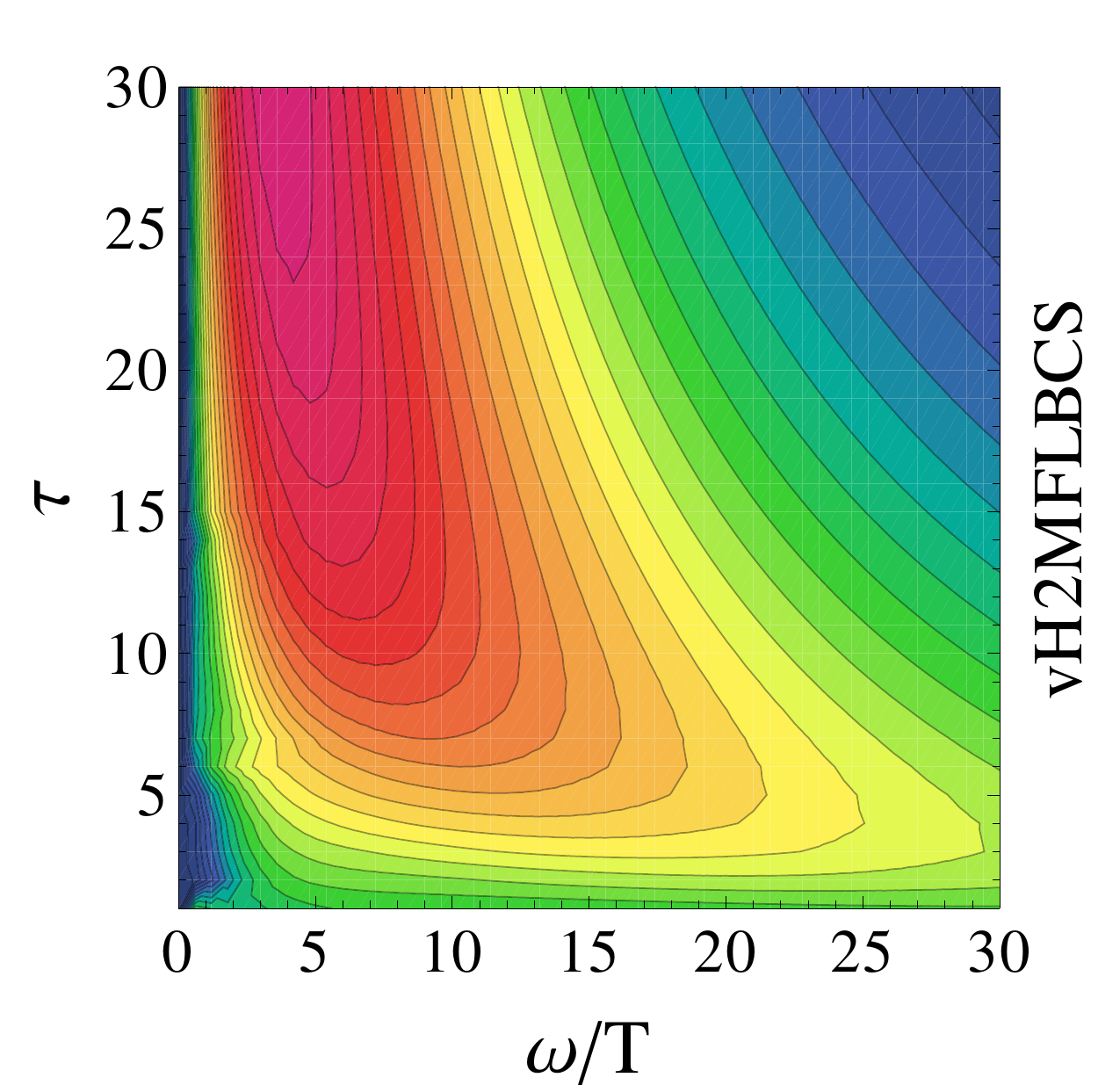}
\caption{(Color online) {\bf Marginal Fermi liquid pair susceptibility with van Hove singularities.} The same plot as Fig.~\ref{fig:MFL1}  for  
  marginal Fermi-liquid with extended van Hove singularities. The pairing interactions are all of the BCS type. 
 The density of states is $N(\epsilon)=|\epsilon|^{-1/2}, |\epsilon+\mu|^{-1/2}, |\epsilon|^{-1/2}\exp(-|\epsilon|/\omega_d)$ for the three different cases respectively. 
Here $T_c=0.01$, and the scaling
  exponent $\delta=1/2$ for all three models. For vH0MFLBCS, the parameters are $a=0.1445, \omega_E=0.05, g=0.2, \omega_b=0.05$. 
For vH1MFLBCS, we take $\mu=-0.25, a=0.2, \omega_E=0.05, g=0.5634, \omega_b=0.05$. For vH2MFLBCS, the parameters are $\omega_d=2, a=0.3, \omega_E=0.4, g=0.3178, \omega_b=0.1$.
 \label{fig:MFL2}}
\end{figure*}

One example is that the van Hove singularity moves away from the Fermi
level (vH1MFLBCS). Consider DOS of the form
$N(\epsilon)=|\epsilon+\mu|^{-1/2}$, we obtain the electronic vertex operator
\begin{equation}
 \Gamma_0(i\nu_n, i\Omega)=\frac{-\pi T}{N_0(i\Omega_+-i\Omega_-)}\left(\frac{1}{\sqrt{i\Omega_++\mu}}-\frac{1}{\sqrt{-i\Omega_+}}\frac{1}{\sqrt{1-i\mu/\Omega_+}}-\frac{1}{\sqrt{i\Omega_-+\mu}}+\frac{1}{\sqrt{-i\Omega_-}}\frac{1}{\sqrt{1-i\mu/\Omega_-}}\right),
\end{equation}
 with $i\Omega_+\equiv i\nu_n+\Omega-\Sigma(i\nu_n+i\Omega)$ and
 $i\Omega_-\equiv -i\nu_n-\Sigma(-i\nu_n)$.
Another example is where there is an extra exponential suppression of
DOS at large energies (vH2MFLBCS),
i.e. $N(\epsilon)=|\epsilon|^{-1/2}\exp(-|\epsilon|/\omega_d)$, for
which one has
\begin{equation}
 \Gamma_0(i\nu_n, i\Omega)=\frac{-\pi T}{N_0(i\Omega_+-i\Omega_-)}(F(\Omega_+)-F(-\Omega_+)-F(\Omega_-)+F(-\Omega_-)),
\end{equation}
with $F(\Omega)=(i\Omega)^{-1/2}\exp(i\Omega/\omega_d){\rm
  Erfc}[(i\Omega/\omega_d)^{1/2}]$. The results are plotted in Fig.~\ref{fig:MFL2}
together with the case where the DOS is of the strict power law
form (vH0MFLBCS). One can see that energy-temperature scaling is broken for the
two deformed cases.

\subsection*{Further comments on the relation between MFL and QCBCS/HS}

The QCBCS/HS approach is not really in conflict with MFL, which is well-known to
be able to capture a large amount of experimental results in cuprates
and heavy fermions. The
pursuit of QCBCS/HS is actually orthogonal to that of MFL. MFL attacks
the single particle Green's functions,
 while QCBCS/HS focuses on the particle-particle channnel. 
Due to vertex corrections, the two channels are not necessarilly simply related. A clear illustration of this is the Luttinger liquid, where these two channels have separate energy-temperature scaling with distinctive exponents.
AdS/CFT seems to provide a natural framework to incorporate such
Luttinger-liquid-type scaling behavior in high dimensional systems,
going well beyond a Hertz-Millis type interpretation of MFL.
 Probing the $AdS_2$ background with fermions, one obtains the MFL type behavior in the fermion Green's functions; 
by probing the $AdS_2$ background with bosonic order parameters, one can
detect energy-temperature scaling in the corresponding susceptibility. 
If we take MFL as synonymous to the fact that the electron scattering rate is proportional to the larger of temperature or frequency,
the contest, that the pair tunneling experiment proposed in this paper is trying to settle, is really between the Hertz-Millis type interpretation
 of MFL and the holographic (or call it the Luttinger-liquid-type) interpretation of MFL.

\end{document}